\documentclass[10pt,journal]{IEEEtran}

\usepackage[noadjust]{cite}

\usepackage{xcolor}
\usepackage{subfigure}
\usepackage{amsmath}
\usepackage{amsthm}
\usepackage{graphicx} 
\usepackage{caption}
\usepackage{graphicx,subfigure}
\usepackage{multicol} 
\usepackage{graphicx} 
\usepackage{amssymb}
\usepackage{float} 
\usepackage{wrapfig} 
\usepackage{booktabs}
\usepackage{lipsum} 
\usepackage{indentfirst}
\usepackage{bm}
\usepackage{color}
\usepackage{amsmath}
\usepackage{enumerate}
\usepackage{extarrows}
\usepackage{url}
\usepackage{diagbox}
\usepackage{graphicx}
\usepackage{epstopdf}
\usepackage{multirow}
\usepackage{tablefootnote}
\usepackage[flushleft]{threeparttable}
\usepackage{slashbox}
\usepackage{array}
\usepackage{amsmath}
\usepackage{enumitem}
\usepackage{framed}
\usepackage{tcolorbox}
\usepackage[ruled,linesnumbered,vlined]{algorithm2e}  
\usepackage{framed}
\usepackage{soul, color, xcolor}

\newtheorem{example}{Example}

\newtheorem{problem}{\textbf{Problem}}

\newtheorem{lemma}{\textbf{Lemma}}

\newtheorem{theorem}{\textbf{Theorem}}

\newtheorem{question}{Question}
\usepackage[colorlinks, linkcolor=blue, anchorcolor=blue, citecolor=blue]{hyperref}

\ifCLASSOPTIONcompsoc
\usepackage[nocompress]{cite}
\else
\usepackage{cite}
\fi

\ifCLASSINFOpdf

\else

\fi

\begin{document}


\title{Link-identified Routing Architecture in Space}

\author{Hefan~Zhang, 
Zhiyuan~Wang, 
Shan~Zhang, 
Qingkai~Meng, 
and Hongbin~Luo
\IEEEcompsocitemizethanks{
\IEEEcompsocthanksitem Part of the results in this paper appeared in WiOpt 2023 \cite{zhang2023link}.
\IEEEcompsocthanksitem This work was supported by National Natural Science Foundation of China under Grants 62225201, 62202021, and 62271019, by National Key Research and Development Program of China under Grant 2022YFB4501000, and by National Engineering Research Center of Advanced Network Technologies under Grant ANT2024003. \textit{(Corresponding author: Zhiyuan Wang)}
\IEEEcompsocthanksitem Hefan~Zhang is with the School of Computer Science and Engineering, Beihang University, Beijing 100191, China. (Email: zhanghefan@buaa.edu.cn).
\IEEEcompsocthanksitem Zhiyuan Wang is with the School of Computer Science and Engineering, Beihang University, Beijing 100191, China, and the State Key Laboratory of Virtual Reality Technology and Systems, Beijing 100191, China. (Email: zhiyuanwang@buaa.edu.cn).
\IEEEcompsocthanksitem Shan Zhang is with the School of Computer Science and Engineering, Beihang University, Beijing 100191, China, and the State Key Laboratory of Software Development Environment, Beijing 100191, China. (Email: zhangshan18@buaa.edu.cn).
\IEEEcompsocthanksitem Qingkai Meng is with the Institute of Artificial Intelligence, Beihang University, Beijing 100191, China. (Email: mengqingkai@buaa.edu.cn).
\IEEEcompsocthanksitem Hongbin Luo is with the School of Cyber Science and Technology, Beihang University, Beijing 100191, China, the School of Computer Science and Engineering, Beihang University, Beijing 100191, China, and the State Key Laboratory of Software Development Environment, Beijing 100191, China. (Email: luohb@buaa.edu.cn).
}
}

\newcommand{\wangRevised}{\color{blue}}
\newcommand{\wangComment}{\color{red}}
\newcommand{\zhangThinkReady}{\bf\color{cyan}}
\newcommand{\response}{\color{brown}}

\markboth{Journal of \LaTeX\ Class Files,~Vol.~14, No.~8, August~2015}%
{Shell \MakeLowercase{\textit{et al.}}: Bare Demo of IEEEtran.cls for Computer Society Journals}

\IEEEtitleabstractindextext{%
\begin{abstract}
Low earth orbit (LEO) satellite networks have the potential to provide low-latency communication with global coverage. 
To unleash this potential, it is crucial to achieve efficient packet delivery. 
In this paper, we propose a Link-identified Routing (LiR) architecture for LEO satellite networks.
The LiR architecture leverages the deterministic neighbor relation of LEO constellations, and identifies each inter-satellite link (ISL).
Moreover, LiR architecture adopts source-route-style forwarding based on in-packet bloom filter (BF).
Each satellite could efficiently encode multiple ISL identifiers via an in-packet BF to specify the end-to-end path for the packets.
Due to false positives caused by BF, the more ISLs are encoded at a time, the more redundant forwarding cases emerge.
Based on the topology characteristics, we derive the expected forwarding overhead in a closed-form and propose the optimal encoding policy.
To accommodate link-state changes in LEO satellite networks, we propose the on-demand rerouting scheme and the on-demand detouring scheme to address the intermittent ISLs.
We also elaborate how to take advantage of LiR architecture to achieve seamless handover for ground-satellite links (GSLs).
Finally, we conduct extensive numerical experiments and packet-level simulations to verify our analytical results and evaluate the performance of the LiR architecture.

\end{abstract}

\begin{IEEEkeywords}
LEO satellite networks, link-identified routing, bloom filter, ISL state management, multicast
\end{IEEEkeywords}}

\maketitle
\def\lenBF{M}
\def\numHash{K}
\def\numElement{N}
\def\packetContentSize{C}

\IEEEdisplaynontitleabstractindextext

\IEEEpeerreviewmaketitle

\section{Introduction}
\subsection{Background and Motivation}
Low-earth-orbit (LEO) satellite networks are promising to provide global Internet broadband services, and have attracted increasing attentions from both industry and academia.
Many companies have filed their ambitious plans and issued proposals for large-scale LEO constellations, e.g., Starlink, Kuiper, Iridium, OneWeb, to name a few.
For example, SpaceX has been deploying a multi-shell Starlink constellation. 
By 2027, the entire constellation is expected to consist of 4088 satellites across five orbital shells.
Different from previous ``bent-pipe'' communication pattern, most LEO constellations will be equipped with inter-satellite links (ISLs) relying on Ka/Ku-band or laser.
In this case, the LEO constellation equipped with ISLs is essentially a large-scale network that provides truly global Internet coverage \cite{chen2024shortest}.

LEO satellite networks have the potential to achieve low-latency content delivery for those Internet services requiring immediacy. 
Different from Geo-stationary orbit satellites (orbiting  at an altitude of 35786 km), the LEO satellites are closer to the Earth, thus can ensure low latency and real-time responsivity. 
This could enable many delay-sensitive Internet services in the future.  
To unleash this potential, it is crucial to achieve efficient data delivery (i.e., routing and forwarding) in LEO satellite networks.

The terrestrial networks are fixed and wired, thus adopt host-centric routing (i.e., TCP/IP).
However, the topology characteristics of LEO satellite networks is different from that of terrestrial networks.
The LEO satellite constellation exhibits a deterministic neighbor relation \cite{wang2024enabling}.
This means that each satellite maintains ISLs with its four neighbor satellites, i.e., two in the same orbit and two in adjacent orbits.
Although these ISLs may be intermittent, the four neighbor satellites are known in advance.
For terrestrial Internet, the neighbor relation is prior unknown.
Hence it is necessary to identify the hosts and form the neighbor relation via message exchanging.
If IP architecture is directly adopted in LEO constellations, then the aforementioned topology characteristics will be ``buried''.
This motivates us to investigate the first key question:

\begin{question}
How to leverage the topology characteristics of LEO satellite networks to achieve efficient data forwarding?
\end{question}

A promising approach is to identify each ISL with a unique identifier instead of naming the satellite nodes or interfaces.
This way, satellites could specify the source-route-style forwarding information (based on the ISL identifiers) into the packet header to guide packet forwarding.
This goal could be achieved via an in-packet bloom filter (e.g., \cite{Lipsin,bloomFilterOptimize4,in-packet,revisit_ip_multicast}).
Similar ideas have been proved efficient in different scenarios (e.g., the publish/subscribe inter-networking~\cite{Lipsin}).
Roughly speaking, bloom filter (BF) is a probabilistic data structure, and can efficiently record multiple elements, but may result in false positives \cite{BloomFilter}.
Such a false positive will lead to redundant forwarding and increase the  forwarding overhead.
Previous studies on BF-based forwarding do not consider how to reduce forwarding overhead due to topology complexity.
The deterministic neighbor relation in LEO satellite networks motivates us to investigate the following key question:
\begin{question}
How to optimize the BF-based forwarding based on topology characteristics of LEO constellation?
\end{question}

Despite the deterministic neighbor relation, LEO constellations also exhibit link-state changes.
On the one hand, inter-satellite links (ISLs) are intermittent due to the high speed relative movement between neighbor satellites.
On the other hand, ground-satellite links (GSLs) are faced with frequently handover due to the short orbital period of LEO satellites.
The two aspects motivate us to consider the third key question:
\begin{question}
How to address the topology dynamics under the source-route-style forwarding in LEO constellation.
\end{question}

To address the three key questions above, we will investigate the topology characteristics of LEO constellations and optimize the BF-based forwarding.
We believe that our study in this paper could lay the underground for efficient packet delivery in LEO constellations.

\subsection{Main Results and Key Contributions}
This paper proposes a Link-identified Routing (LiR) architecture for LEO satellite networks.
Specifically, the LiR architecture identifies the ISLs and adopts the source-route-style forwarding based on an in-packet BF.
Under the LiR architecture, each satellite could encode multiple ISL identifiers via an in-packet BF to specify the end-to-end path of the packets.
Due to unavoidable false positives caused by the BF, the more ISLs encoded at a time, the more redundant forwarding will emerge.
To reduce the forwarding overhead, we investigate the optimal segment encoding policy design, seeking the right balance between encoding delay and forwarding overhead.
Our key contributions are as follows:
\begin{itemize}
\item \textit{A Novel Routing Architecture for LEO Constellations:}
We propose a source-route-style LiR architecture for LEO constellations that eliminates the need for end-to-end addressing and enables efficient path control and data forwarding.
Specifically, this architecture allows satellites to specify the path by encoding ISL identifiers into the in-packet BF.
Moreover, We analytically derive the expected forwarding overhead of this architecture based on the topology characteristics of LEO constellations and the false positive rate of the BF.
The analytical results facilitate our later optimization for the LiR architecture.

\item \textit{Segment Encoding Policy Design:}
We introduce the segment encoding policy design, allowing the source and intermediate satellites to jointly encode part of ISLs towards the destination.
We first characterize the segment encoding policy in a unified framework, and then formulate the optimal segment encoding problem as a binary non-linear programming.
Based on the decomposable structure, we propose an algorithm to solve it efficiently.

\item \textit{Link-State Management:}
We design link-state management schemes for the LiR architecture to address the occasional ISL failures and frequent GSL handovers.
First, we devise the on-demand rerouting and detouring schemes for the LiR architecture to address the intermittent ISLs.
Second, we elaborate how to leverage the LiR architecture to achieve seamless GSL handover via multicast transmission.

\item \textit{Packet-Level Performance Evaluation:} 
We evaluate the performance of our proposed LiR architecture via packet-level experiments under Iridium constellation on OMNeT++.
The experiments validate our analytical results, and demonstrate the performance of LiR architecture in three aspects. 
First, the optimal segment encoding policy significantly reduces the queuing delay compared to source encoding.
Second, our proposed on-demand rerouting and detouring schemes outperform the classic link state announcement scheme in terms of addressing occasional ISL failures.
Third, the multicast transmission under the LiR architecture significantly benefits the one-to-many traffic pattern to achieve seamless GSL handover.
\end{itemize}

The rest of this paper is as follows.
Section~\ref{Section: Literature Review} reviews related literature.
Section~\ref{Section: ISL-Identified forwarding framework} introduces our proposed LiR architecture.
Section~\ref{Section: Encoding Scheme Optimization} presents the encoding policy design.
We introduce how to address occasional ISL failures and frequent GSL handovers under the LiR architecture in
Section~\ref{Section: ISL Failure Management} and Section~\ref{Section: Multicast Configuration}, respectively.
Section~\ref{Section: Experimental Results} provides packet-level simulation results.
Section~\ref{Section: Conclusion and Future Works} concludes this paper.

\section{Literature Review}
\label{Section: Literature Review}

There have been many studies on LEO satellite networking, which focus on intra-domain routing (e.g., host-centric LoFi~\cite{shan2023routing}, content-centric LPIH~\cite{yan2024logic}, comparison~\cite{yan2023comparative}), inter-domain routing (e.g., host-centric BGP-S~\cite{ekici2001network} and content-centric PCR~\cite{zeng2024adaptive}), congestion control (e.g., \cite{huang2024how}\cite{cao2023satcp}), and emulation platforms (e.g., LeoEM~\cite{cao2023satcp}, StarryNet~\cite{lai2023starrynet}, and OpenSN~\cite{lu2024opensn}).
This paper presents the Link-identified Routing (LiR) architecture.
We will review two streams of literature that are mostly related to our study in this paper.


\subsection{Source-Route-Style Forwarding in Satellite Networks}
LEO constellations have several advantages (e.g., low latency, global coverage, and high bandwidth), making it promising for mission-critical data transmission.
However, satellites in open environments are threatened by various types of attacks (e.g., Signal Interference, Eavesdropping, and DDoS) \cite{cao2020analysis}.
Moreover, the mobility of satellites frequently expose them to many untrusted areas (e.g., hostile country).
Some studies (e.g., \cite{goetzelmann2012space}) suggest that mission-critical data should avoid these areas to reduce security risks.
Since source routing allows the source satellite to schedule its path to the destination, it could easily fulfill
that requirement.
Regarding source routing, it has been widely studied in different networks.
Sunshine in \cite{Source_Routing} provides a detailed introduction as well as its advantages and weakness.
Additionally, a few studies (e.g. \cite{Satellite_Network_Source_Routing, DSRSC, CEMR, LEO-MEO}) investigate the feasibility of source routing in LEO satellite networks, which demonstrates its potential in achieving low control overhead, multipath routing, and efficient packet forwarding.
However, these studies overlook how to efficiently record the end-to-end path and conduct data forwarding.
Different from these studies, our proposed LiR utilizes a space-efficient data structure (i.e., in-packet bloom filter) to achieve efficient path control and data forwarding.


\begin{figure*}
\centering
\subfigure[ISL identification]{\label{fig: link-identifier}
\includegraphics[height=0.22\linewidth]{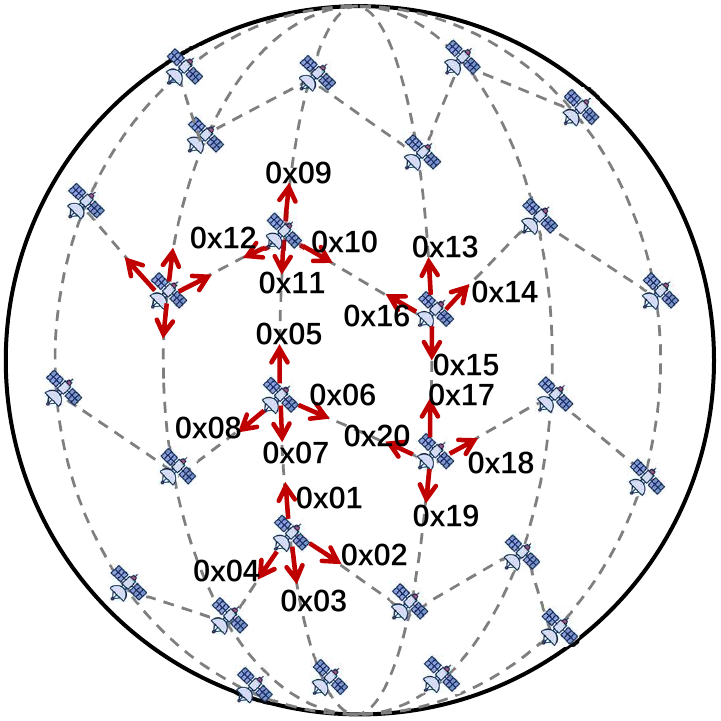}}\quad
\subfigure[Illustration of BF]{\label{fig: bloomfilter}
\includegraphics[height=0.22\linewidth]{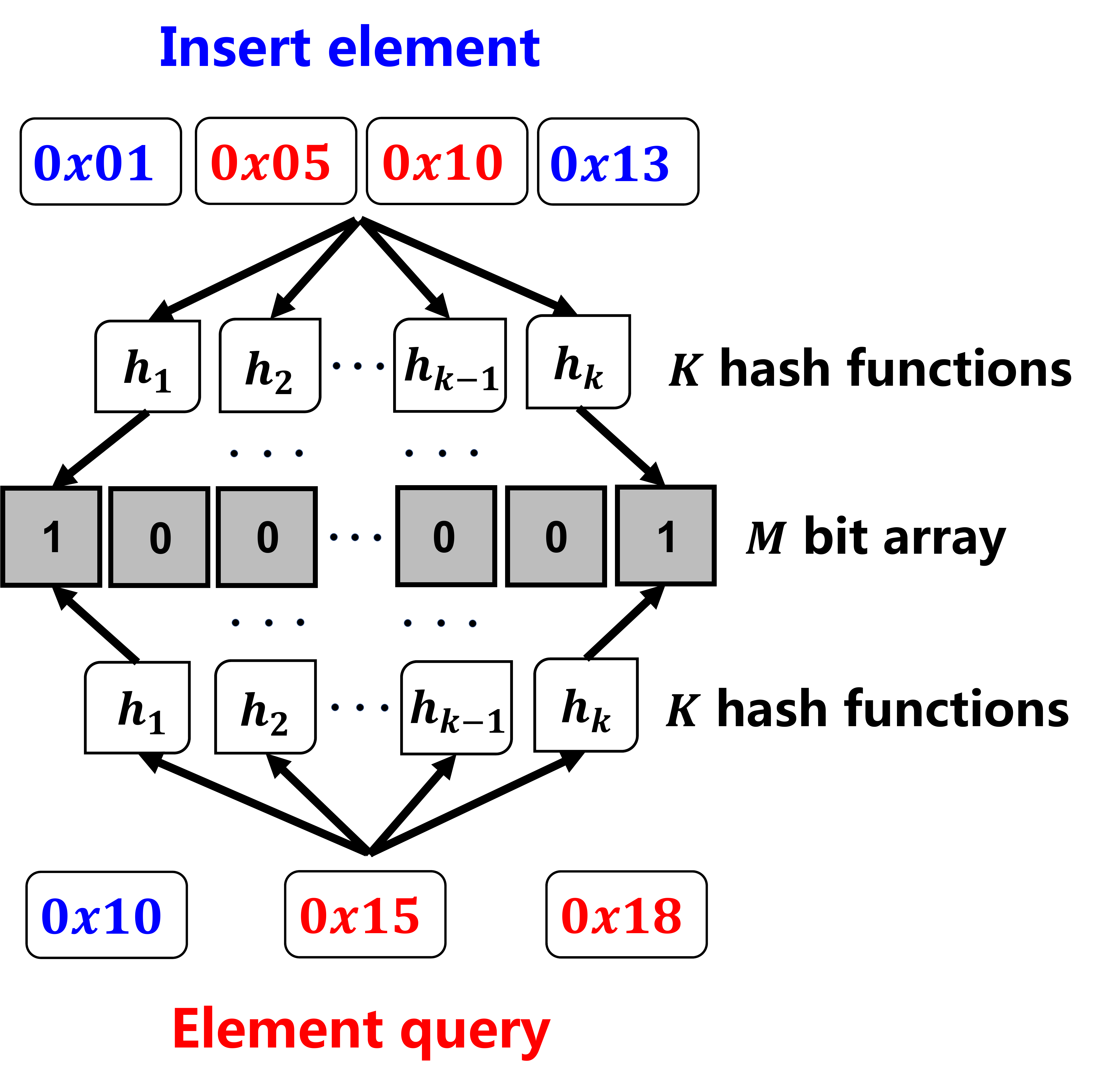}}\quad
\subfigure[BF-based forwarding]{\label{fig: Re-EncodeExample}
\includegraphics[height=0.22\linewidth]{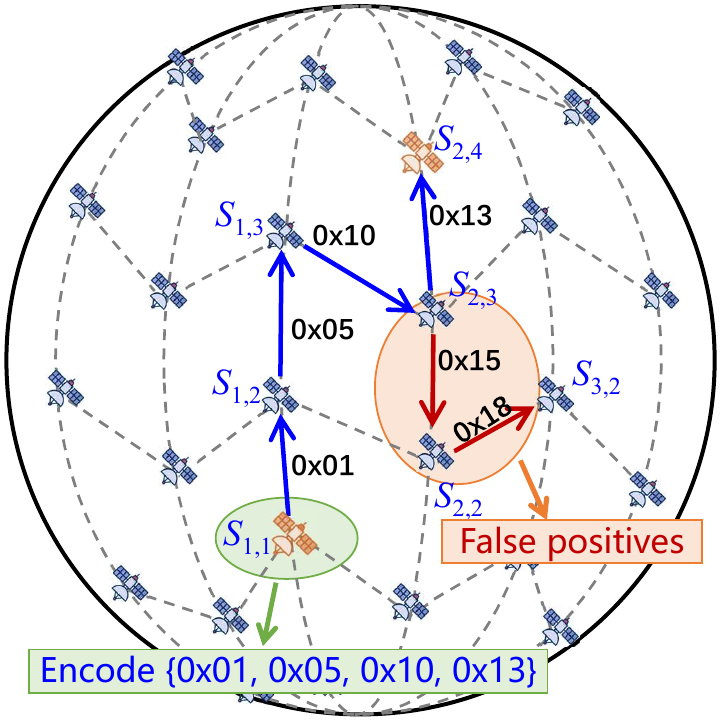}}\quad
\subfigure[Segment encoding]{\label{fig: segment-encoding-vis}
\includegraphics[height=0.22\linewidth]{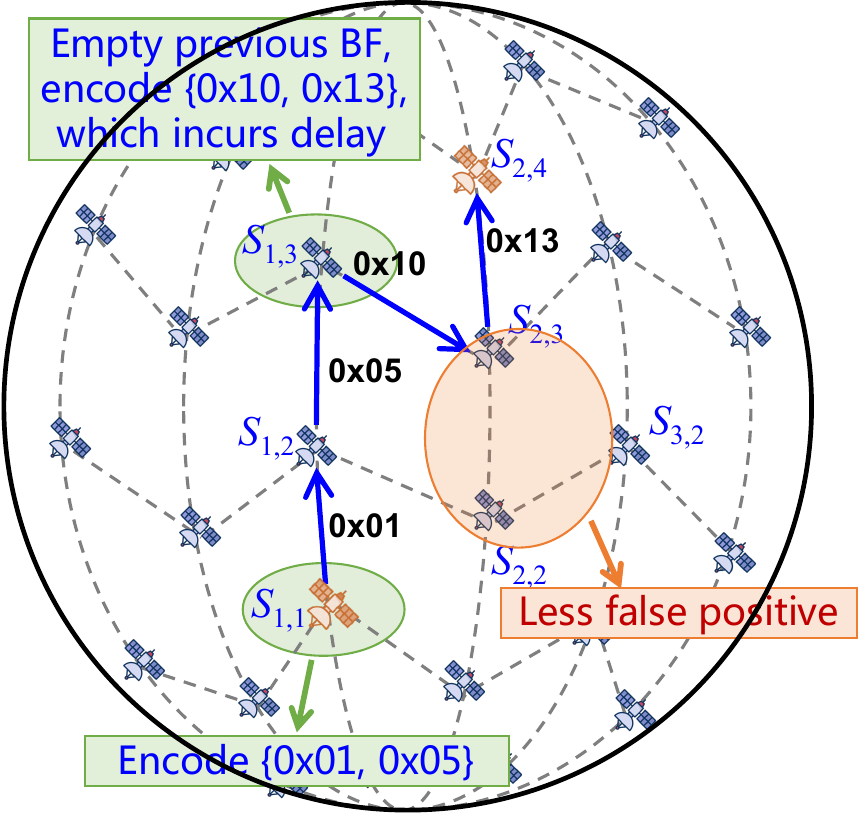}}
\caption{Illustration of BF-based forwarding under LiR}
\end{figure*}

\subsection{BF-based Forwarding}

There have been many studies on BF-based forwarding in different networks (see \cite{BloomFilter} for a comprehensive survey).
Overall, BF has two working modes in networking scenarios. 
\begin{itemize}
\item The \textit{in-router} BF functions similarly to routing tables in IP networks (e.g., \cite{longest_prefix_match}) and forwarding information bases in named-data networks (e.g., \cite{ndn1,ndn2,ndn3}).
Multiple entries can be compressed and quickly checked.

\item The \textit{in-packet} BF provides a flexible and efficient way to implement source-route-style forwarding.
Petri \textit{et al.} in \cite{Lipsin} propose a multicast forwarding fabric called LIPSIN, which applies the BF in the packet to store link identifiers of the multicast tree. 
Tapolcai \textit{et al.} in \cite{bloomFilterOptimize4} propose a multistage BF to reduce the false positive rate and improve the scalability of LIPSIN. 
Rothenberg \textit{et al.} in \cite{data-center} propose a networking approach for cloud data centers based on in-packet BF. 
We refer interested readers to \cite{in-packet} for more usages of the in-packet BF. 
\end{itemize}

\textit{Compared to these studies, our proposed LiR architecture makes two key improvements}.
First, we take into account the topology characteristics of LEO constellations and derive the expected forwarding overhead.
Second, we propose the optimal segment encoding design, seeking for a proper balance between forwarding overhead and encoding delay.
The two aspects are especially crucial to achieve efficient data delivery in LEO satellite networks with limited bandwidth.
Furthermore, compared to our previous study~\cite{zhang2023link}, this paper further devises the link-state management scheme for the LiR architecture.

\section{LiR Architecture for LEO Constellations}
\label{Section: ISL-Identified forwarding framework}
This section introduces the Link-identified Routing (LiR) architecture for LEO satellite networks and evaluates its performance.
We first introduce ISL identification in Section~\ref{Section: Satellite Network Architecture}, followed by the in-packet BF based forwarding procedure in Section~\ref{Subsection: ISL-Identified Forwarding based on BF}.
In Section~\ref{Subsection: Forwarding Performance Analysis}, we analyze the forwarding performance.
Section~\ref{Section: Comparison Between LiR and SRv6} compares LiR with SRv6.
Finally, we evaluate the path representation efficiency of LiR in Section~\ref{Section: evaluation of path representation efficiency}.

\subsection{ISL Identification for LEO Constellation}
\label{Section: Satellite Network Architecture}
The LEO constellation is composed of multiple orbits of satellites.
Each satellite connects to its four neighbor satellites in typical polar constellations (e.g., Iridium and OneWeb) via wireless ISLs.
Such a neighbor relation and the ISLs are predetermined and fixed when the constellation is deployed.
This feature allows us to identify the ISLs (instead of naming the interfaces) and adopt source-route-style forwarding based on the ISL identifiers.

The identification of ISLs can be implemented in different ways, but the ISL identifiers needs to be globally unique.
For example, the interface MAC address of each satellite is one feasible proposal for ISL identification. 
In our proposed LiR architecture, we assign each \textit{unidirectional} ISL a globally unique identifier. 
In this case, each satellite is associated with four ISL identifiers.
Fig.~\ref{fig: link-identifier} illustrates ISL identification, where each ISL direction is uniquely identified.

Given the ISL identification, we can construct source-route-style information to guide the packet forwarding at intermediate nodes.
To this end, we will follow previous studies (e.g.,~\cite{Lipsin,bloomFilterOptimize4,in-packet,revisit_ip_multicast}) and take advantage of in-packet BFs.


\subsection{BF-based Packet Forwarding}
\label{Subsection: ISL-Identified Forwarding based on BF}
BF is a space-efficient probabilistic data structure (i.e., a binary vector), which enables a constant-time membership query.
It has been widely used to check whether an element is a member of a set.
With a BF in the packet header, it is efficient to encode the source-route-style forwarding information (i.e., the ISL identifiers).
This allows the intermediate satellites on the route to determine which outgoing ISLs the packet should be forwarded to.
Furthermore, BF may result in false positives, while no false negatives.
That is, the outcome of an element query based on the BF is either ``\textit{possibly in the set}'' or ``\textit{definitely not in the set}''.
This means that the recorded ISLs will not be missed in the forwarding process under LiR.
LiR may lead to redundant forwarding, but the correct forwarding direction will not be affected by the BF property.
Therefore, LiR adopts BF-based forwarding, but will not cause packet loss along the path recorded in the packet.

\subsubsection{Bloom Filter (BF)} 
BF is formally defined as an $\lenBF$-bit binary vector and $\numHash$ hash functions, which will be used to represent $\numElement$ elements (i.e., ISL identifiers).
We let $h_k(\cdot)$ denote the $k$-th hash function. 
Each hash function takes the element 
(to be represented) as the input, and randomly maps the element to one of the positions in the $\lenBF$-bit vector.
We introduce the element insertion and query based on Fig.~\ref{fig: bloomfilter}.
\begin{itemize}
\item \textit{Element Insertion:}
The $\lenBF$-bit vector of a BF is initialized to be all-zero.
If one needs to insert an element $x$, then the $\numHash$ independent hash functions will be employed, mapping the element $x$ to $\numHash$ positions in the $\lenBF$-bit vector. 
Accordingly, the $\numHash$ positions of the bit vector will be set to 1.
Note that multiple elements could be inserted into a BF, which may cause false positives {during queries}.

\item \textit{Element Query:} 
To query whether an element is encoded into the BF, we will feed it to the $K$ hash functions to obtain the $K$ array positions.
If any bit at these positions is 0, then the element is definitely not in the BF. 
Otherwise, if all bits at the $K$ positions are 1, then either the element is in the set, or these bits were set to 1 by chance during the insertion of other elements.
Previous studies (e.g., \cite{BloomFilter}) have shown that the false positive rate of representing $N$ elements via an $\lenBF$-bit BF under $\numHash$ hash functions is given by
\begin{equation}\label{Equ: p}
\begin{aligned}
    p(\lenBF,\numElement,\numHash) &=
    \left[1 - \left(1 - \frac{1}{\lenBF}\right)^{\numHash\numElement}\right]^{\numHash}.
\end{aligned}    
\end{equation}
\end{itemize}

With the BF in the packet header, the satellite can store a variable number of ISL identifiers in the packet without taking up much space.
This makes the source-route-style forwarding feasible in satellite networks.
Next, we introduce BF-based forwarding process in details, and then  analyze the forwarding performance caused by false positives in Section~\ref{Subsection: Forwarding Performance Analysis}.

\subsubsection{BF-based Forwarding}
We introduce the BF-based packet forwarding process based on the example in Fig.~\ref{fig: Re-EncodeExample}.
Consider the data delivery from satellite $\textit{S}_{1,1}$ to satellite $\textit{S}_{2,4}$.
The detailed forwarding process is as follows:
\begin{itemize}
\item \textbf{Path Calculation and Encoding:} 
The source satellite $\textit{S}_{1,1}$ calculates the path towards the destination satellite $\textit{S}_{2,4}$ via Dijkstra algorithm, and and obtains the path represented by four ISL identifiers \{0x01, 0x05, 0x10, 0x13\}.
The source satellite $\textit{S}_{1,1}$ encodes the four ISL identifiers into the BF of the packets (to be delivered).

\item \textbf{Data Forwarding:}
Upon receiving the packets, intermediate satellites determine the forwarding direction by checking whether the other three outgoing ISLs (except for the incoming ISL) are recorded in the in-packet BF.
For example, when $\textit{S}_{1,3}$ receives a packet from $\textit{S}_{1,2}$, it finds that link identifier 0x10 has been encoded in the BF.
Accordingly, $\textit{S}_{1,3}$ will forward the packet through the outgoing ISLs associated with link identifier 0x10.
Following this logic, the packet ultimately reaches its destination (i.e., $\textit{S}_{2,4}$).
\end{itemize}

During the forwarding process above, false positives will result in incorrect forwarding, which occurs with a probability $p(\lenBF,\numElement,\numHash)$ defined in (\ref{Equ: p}).
In this case, the packet will be forwarded to the ISL that is not along the path between the source and destination satellites. 
More severely, such an incorrect forwarding may continue, thus wastes the limited ISL bandwidth and causes queuing delay for other packets. 
The red arrows in Fig.~\ref{fig: Re-EncodeExample} represent a two-hop incorrect forwarding. 
Next, we analyze the performance of BF-based forwarding.

\subsection{Forwarding Performance Analysis}
\label{Subsection: Forwarding Performance Analysis}
We analyze the overhead caused by BF-based forwarding, including incorrect and correct forwarding overhead.
\begin{itemize}
\item \textit{Incorrect Forwarding Overhead:}
Recall that BF is a probabilistic data structure, indicating that an element is either definitely not in the set or possibly in the set.
Hence false positives will lead to incorrect forwardings towards unspecified ISLs.
We let $f_{\textit{IFO}}(\cdot)$ denote the incorrect forwarding overhead, which measures the data volume (in KB) delivered on the incorrect ISLs.

\item \textit{Correct Forwarding Overhead:}
The BF-based forwarding utilizes the $\lenBF$-bit vector to record the ISL identifiers, which also increases the forward overhead along the correct path.
We let $f_{\textit{CFO}}(\cdot)$ denote the correct forwarding overhead, which measures the data volume (in KB) caused by the $\lenBF$-bit BF along the planned route.
\end{itemize}
These two types of forwarding overhead jointly contribute to the negative impact of the BF-based forwarding on the satellite network.
Ideally, we would like to reduce them as much as possible.
To have a better understanding, we first derive the forwarding overhead in a closed-form.
Theorem~\ref{Theorem: incorrect} presents the closed-form expression for incorrect forwarding overhead.

\begin{theorem}\label{Theorem: incorrect}
Given a route with $\numElement$ ISL identifiers encoded into an $\lenBF$-bit BF under $\numHash$ independent hash functions, the expected incorrect forwarding overhead is given by
\begin{equation}
\begin{aligned}
    f_{\textit{IFO}}(\numElement,\lenBF,\numHash) 
    &=\frac{(2\numElement+1)(\lenBF+\packetContentSize) p(\numElement,\lenBF,\numHash)}{1-3p(\numElement,\lenBF,\numHash)},
\end{aligned}
\end{equation}    
where $p(\numElement,\lenBF,\numHash)$ denotes the false positive rate defined in (\ref{Equ: p}), and $\packetContentSize$ denotes the volume of effective data in the packet.
\end{theorem}

\begin{figure}
\centering
\includegraphics[width=\linewidth]{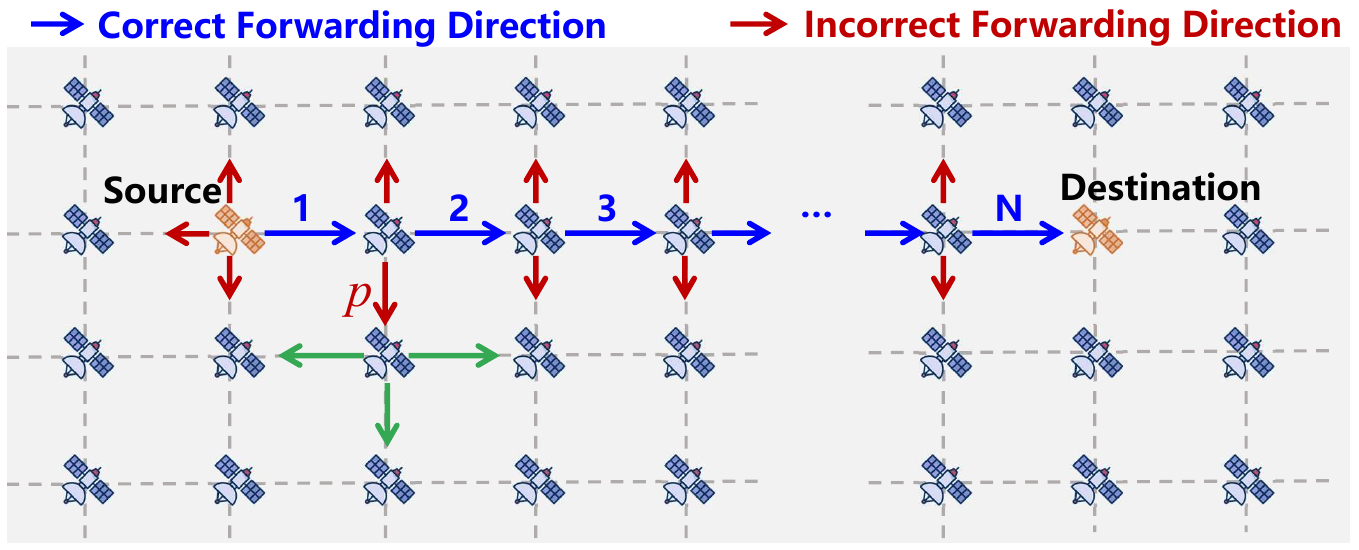}
\caption{Illustration of incorrect packet forwarding}
\label{fig: packetForwardingProcess}
\end{figure}
\begin{proof}[\bf Proof of Theorem \ref{Theorem: incorrect}]
We prove this theorem by calculating the expected incorrect forwarding overhead.
Specifically, $\lenBF+\packetContentSize$ represents the per-hop incorrect forwarding overhead.
Moreover, given an $\numElement$-hop path, there are $2\numElement+1$ potential incorrect forwarding directions (as shown by the red arrows in Fig.~\ref{fig: packetForwardingProcess}).    
Given the false positive rate $p$, we let $E(p)$ denote the expected number of forwarding hops towards a single incorrect forwarding direction.
Hence we have 
\begin{equation}\label{Proof Equ: f_IFO}
f_{\textit{IFO}}(N,\lenBF,\numHash)
= 
(2\numElement+1)(\lenBF+\packetContentSize)E(p).
\end{equation}
Now we derive the function $E(p)$ in a recursive manner, considering the following two cases.
\begin{itemize}
    \item Along an incorrect forwarding direction, the incorrect forwarding event does not occur with the probability $1-p$.
    In this case, the number of forwarding hops is $0$.
    \item Along an incorrect forwarding direction, the incorrect forwarding event occurs with the probability $p$.
    In this case, the number of forwarding hops is $1+3E(p)$, since there are three more incorrect forwarding directions.
\end{itemize}
The above discussion leads to the following equation
\begin{equation}
    E(p)=(1-p)\cdot0+p\cdot\left[1+3E(p)\right].
\end{equation}
Solving the above equation with respect to $E(p)$ yields
\begin{equation}\label{Proof Equ: H(p)}
    E(p)=\frac{p}{1-3p}.
\end{equation}
Substituting (\ref{Proof Equ: H(p)}) into (\ref{Proof Equ: f_IFO}) completes the proof.    
\end{proof}


Theorem \ref{Theorem: incorrect} indicates that a larger false positive rate $p(N,\lenBF,\numHash)$ will increase the incorrect forwarding overhead.
If the source satellite encodes a lot of ISL identifiers (i.e., $N$ is large), then the incorrect forwarding overhead $f_{\textit{IFO}}(\cdot)$ will be definitely great.
To address this issue, we need to consider \textit{segment encoding} (to be discussed in Section~\ref{Section: Encoding Scheme Optimization}).
Additionally, if the source satellite uses a large BF (i.e., $\lenBF$ is large), then the incorrect forwarding overhead $f_{\textit{IFO}}(\cdot)$ will be small.
However, this will increase the correct forwarding overhead $f_{\textit{CFO}}(\cdot)$, which is presented in Theorem~\ref{Theorem: correct}.

\begin{theorem}\label{Theorem: correct}
Given a route with $\numElement$ ISL identifiers encoded into an $\lenBF$-bit BF, then the correct forwarding overhead is 
\begin{equation}
    f_{\textit{CFO}}(\numElement,\lenBF) 
    = \lenBF \numElement.
\end{equation}
\end{theorem}

Based on the results in Theorem~\ref{Theorem: incorrect} and Theorem~\ref{Theorem: correct}, we obtain the total forwarding overhead as follows:
\begin{equation}\label{Equ: Forwarding_Overhead}
\begin{aligned}
f_{\textit{FO}}(\numElement,\lenBF,\numHash) 
=& f_{\textit{IFO}}(\numElement,\lenBF,\numHash)  + f_{\textit{CFO}}(\numElement,\lenBF).
\end{aligned}    
\end{equation}

Based on (\ref{Equ: Forwarding_Overhead}), one could properly design the BF to reduce the forwarding overhead.
First, the number of hash functions $\numHash$ is usually fixed at the stage of network initialization.
Hence we will view $\numHash$ as a constant in this paper.
Second, the length of BF $\lenBF$ plays a significant role on the correct and incorrect forwarding overhead.
Ideally, we aim to adopt the optimal BF size given the forwarding distance (measured by the number of ISL identifiers $N$).
Mathematically, we let $f(\numElement)$ denote the forwarding overhead incurred by an $\numElement$-hop delivery under the optimal BF length.
That is, $f(\cdot)$ is defined as follows:
\begin{equation}\label{Equ: F(N)}
\begin{aligned}
    f(\numElement)
    \triangleq
    \min\limits_{\lenBF\ge0} &\quad f_{\textit{FO}}(\numElement,\lenBF,\numHash).
\end{aligned}
\end{equation}



\begin{figure}
\centering
\includegraphics[width=0.55\linewidth]{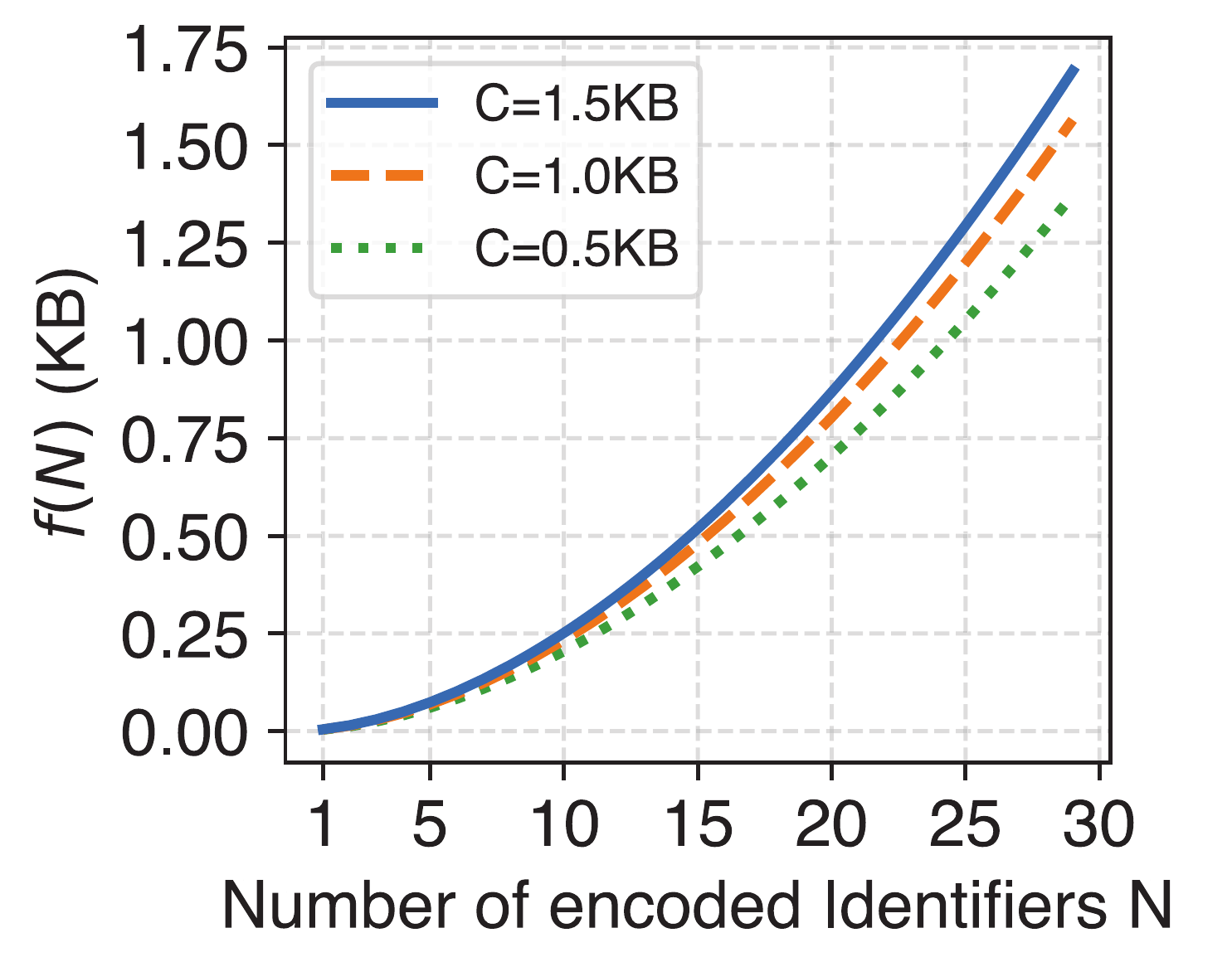}
\caption{$f(N)$ given $\numHash=5$}
\label{fig: k=5}
\end{figure}

It is not difficult to derive (\ref{Equ: F(N)}), as it is a one-dimensional optimization.
Nevertheless, $f(N)$ is still positively related to the number of encoded ISL identifiers $N$.
Fig.~\ref{fig: k=5} shows how $f(N)$ increases with $N$ when the number of hash functions is $\numHash=5$.
Note that $f(N)$ is convexly increasing with the number of encoded ISL identifiers $N$.
That is, the marginal forwarding overhead is increasing.
This means that it is not beneficial to encode too many ISL identifiers at a time, even though we have adopted the optimal BF length.
This observation motivates us to further investigate how to encode the ISL identifiers (from source satellite to destination satellite) segmentally via intermediate relay satellites.
We will present the optimal design for the \textit{segment encoding} in Section~\ref{Section: Encoding Scheme Optimization}.
Before that, we introduce the main difference between the LiR architecture and segment routing over IPv6 (SRv6).

\subsection{Comparison Between LiR and SRv6}
\label{Section: Comparison Between LiR and SRv6}
Both LiR and SRv6 allow source nodes to specify the end-to-end path in the packet header.
However, they are different in terms of the header structure and the forwarding process.
Next we briefly introduce how they work and the main differences.

\begin{figure}
\centering    
\includegraphics[width=0.9\linewidth]{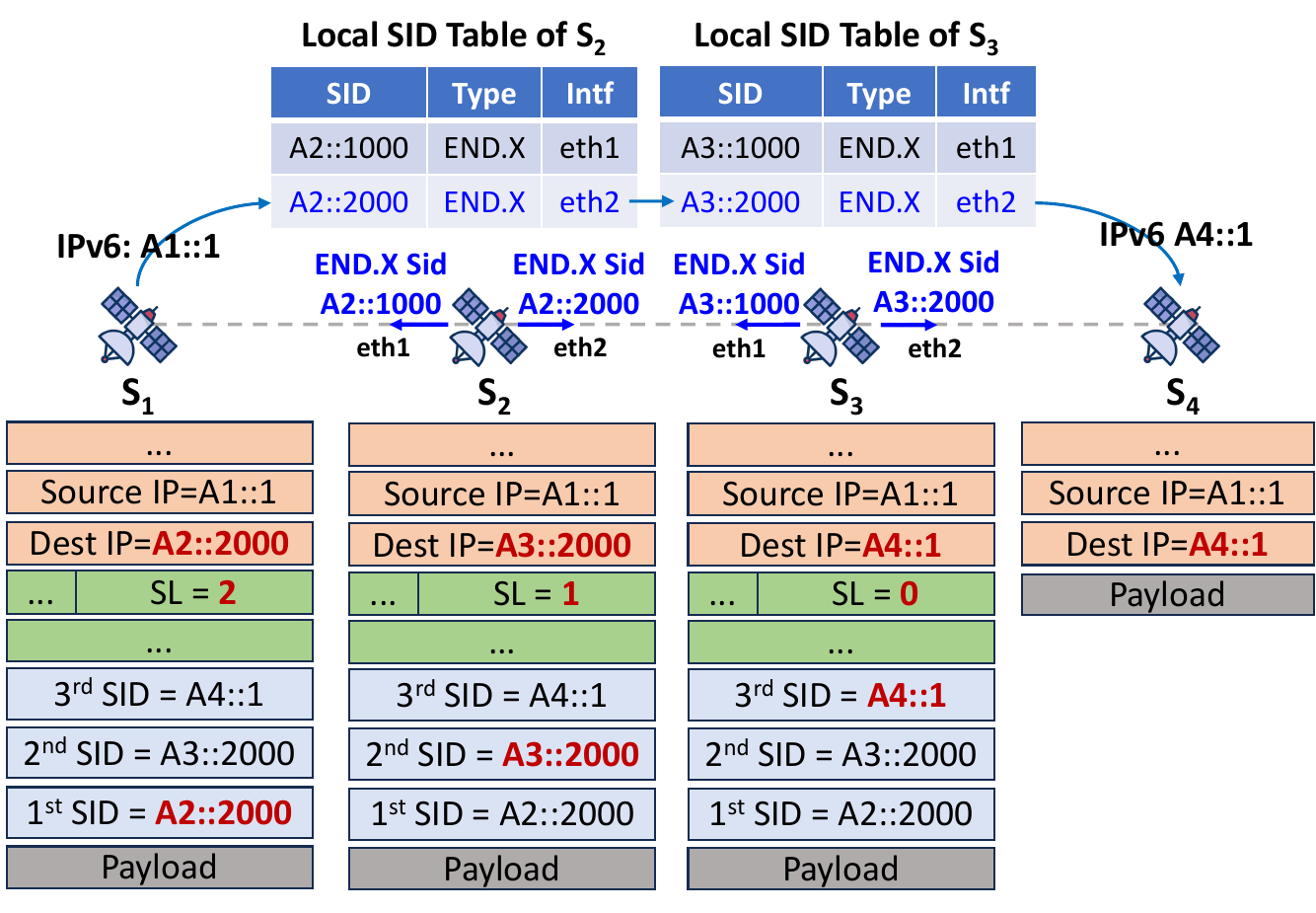}
\caption{Illustration of SRv6.}
\label{fig: srv6 example}
\end{figure}

\begin{figure*}
\centering
\subfigure[Payload ratio of packets]{\label{fig: lir sr ratio}
\includegraphics[width=0.46\linewidth]{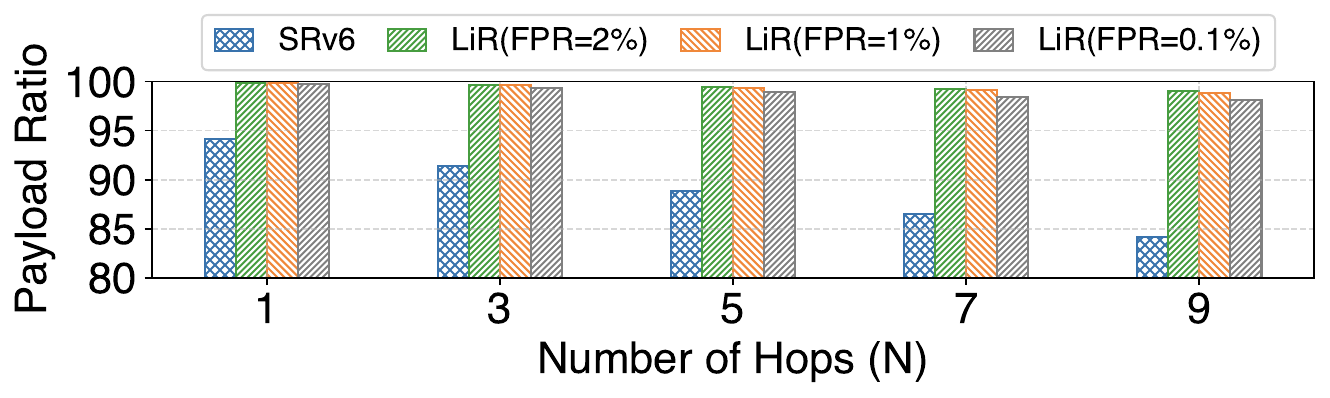}}\qquad
\subfigure[Forwarding overhead]{\label{fig: lir sr forwarding overhead}
\includegraphics[width=0.46\linewidth]{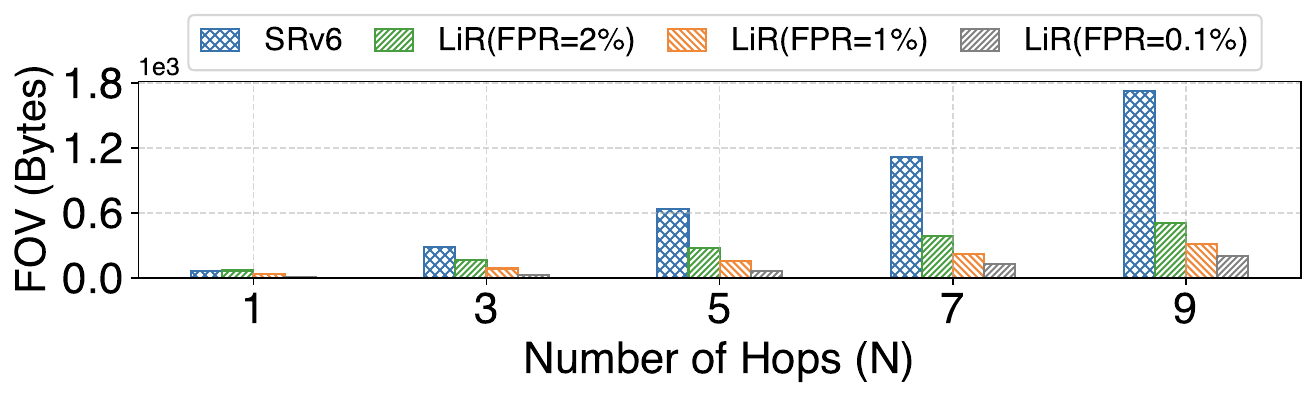}}
\caption{Numerical results of LiR and SRv6}
\label{fig: SR LIR Numerical Experiments}
\end{figure*}

\subsubsection{How SRv6 Works}
Segment Routing over IPv6 (SRv6) is a protocol based on the IPv6 forwarding plane.
SRv6 specifies explicit path using 128-bit Segment Identifiers (SIDs) within the Segment Routing Header (SRH).
As shown in Fig.~\ref{fig: srv6 example}, satellite $\textit{S}_{1}$ inserts SIDs (i.e., A2::2000, A3::2000, and A4::1) into the segment list of the SRH.
There is a field called Segments Left (SL) in SRH, which indicates the number of intermediate nodes to be visited before reaching the destination.
Satellite $\textit{S}_{1}$ sets this value to 2, since there are two intermediate nodes (i.e., satellites $\textit{S}_{2}$ and $\textit{S}_{3}$) between it and the destination satellite $\textit{S}_{4}$.
When an intermediate node receives a packet, it checks the destination IPv6 address in the local SID table to determine the appropriate action.
Different types of SIDs corresponds to different actions.
As shown in Fig.~\ref{fig: srv6 example}, upon receiving the packet from satellite $\textit{S}_{1}$, satellite $\textit{S}_{2}$ will lookup its local SID table for A2::2000, the corresponding actions of this entry will be adopted: the SL field is decreased by one, and the destination IP address is updated to the next SID (i.e., A3::2000).

\subsubsection{How LiR Works} 
LiR is a network layer architecture, which specifies the end-to-end path via the in-packet BF.
BF allows LiR to efficiently record multiple elements.
When an intermediate satellite receives an LiR packet, it checks whether the outgoing link identifiers are encoded in the packet and then forwards the packet accordingly.
Due to false positive, the packet may be forwarded to an unspecified link. 

\subsubsection{Main Differences}
The main differences between SRv6 and LiR lie in the data structures used to specify the path.
This difference renders the forwarding process substantially different, and also causes different forwarding overhead.
\begin{itemize}

\item \textbf{Efficiency of Path Control:} In SRv6, the source node utilizes a deterministic data structure to specify the end-to-end path.
Compared to SRv6, LiR leverages a more efficient data structure (i.e., in-packet BF) to record the forwarding path, which provides better scalability and bandwidth utilization.
However, the false positives of in-packet BF may lead to incorrect forwardings and affect the normal transmission of other packets.
To mitigate this, we introduce \textit{segment encoding} (to be discussed in Section~\ref{Section: Encoding Scheme Optimization}).
This allows the source and some intermediate satellites to collaboratively encode ISLs towards the destination, which reduces incorrect forwardings.

\item \textbf{Efficiency of Data Forwarding:} In SRv6, each intermediate node has to process SRH (to determine the next segment), update the destination IP, and lookup the routing table to determine the next hop.
As mentioned in \cite{bolla2010energy}, routing table lookups consume about 32\% of the total power consumption of IP routers.
As the constellation scales, both the SRH size and the routing table grow rapidly, leading to increased time and energy overhead.
In contrast, under LiR, each intermediate node only needs to check whether its three outgoing ISLs (except for the incoming ISL) are encoded in the in-packet BF, and forward the packets accordingly.
As a result, LiR's simpler forwarding process consumes less power and enables faster packet forwarding.
\end{itemize}

The two sub-figures in Fig.~\ref{fig: SR LIR Numerical Experiments} compare LiR and SRv6 in terms of payload ratio and forwarding overhead, respectively.
In each sub-figure, the horizontal axis represents the number of hops.
The blue bars correspond to SRv6, while the other three bars demonstrate the performance of LiR under different false positive rates.
The payload of each packet is set to 1KB.
Fig.~\ref{fig: lir sr ratio} shows the payload ratio of SRv6 and LiR packets.
LiR achieves a higher payload ratio than SRv6.
This is because LiR relies on BF to record the end-to-end path, which is more efficient than SRv6.
Fig.~\ref{fig: lir sr forwarding overhead} shows the forwarding overhead incurred by SRv6 and LiR.
Note that a higher false positive rate results in higher forwarding overhead for LiR.
Nevertheless, LiR still outperforms SRv6 in terms of reducing forwarding overhead.
Next, we will evaluate the path representation efficiency of LiR by comparing it to other advanced path representation data structures.

\subsection{Evaluation of Path Representation Efficiency} \label{Section: evaluation of path representation efficiency}
As shown in Section~\ref{Section: Comparison Between LiR and SRv6}, the large segment routing header (SRH) in SRv6 leads to higher forwarding overhead compared to LiR.
Cheng~\textit{et al.} in \cite{cheng2020shorter} analyze the overhead of SRH and raise several scalability related concerns (e.g., exceeding path MTU, low payload ratio, and increased processing load).
To address these concerns, many studies aim to optimize path representation.
In the following, we consider two methods of representing the end-to-end path.

\begin{itemize}
\item Explicit Link Representation (ELR) records the end-to-end path via a list of link identifiers.
SRv6 and its variants (e.g., MicroSID~\cite{microSID} and GSRv6~\cite{cheng2020g}) fall into this case.
We investigate the overhead lower bound of these methods.
Specifically, suppose that an LEO constellation has a total of $L$ ISLs, then ELR will use at least $\lceil\log_{2}(L)\rceil$ bits to represent these ISLs.
If the end-to-end path consists of $N$ ISLs, then the length of the packet header is $N\lceil\log_{2}(L)\rceil$ bits.
Accordingly, the total forwarding overhead will be $N^2\lceil\log_{2}(L)\rceil$.
In practice, the overhead incurred by SRv6 and its variants (e.g., MicroSID~\cite{microSID} and GSRv6~\cite{cheng2020g}) is greater than $N^2\lceil\log_{2}(L)\rceil$.

\item Our proposed LiR records the end-to-end path via the in-packet BF.
Due to the false positives, the forwarding overhead of LiR depends on the packet size.
Sarrar~\textit{et al.} in \cite{sarrar2012investigating} analyze the packet size distributions of IPv4 and IPv6 traffic, which reveals that over 82\% percent of IPv6 packets are smaller than 72 bytes.
Many packets have a size of 1280 bytes.
Therefore, we evaluate the forwarding overhead of LiR under these two common packet sizes.
\end{itemize}

\begin{figure}
\centering
\subfigure[Packet header size]{\label{fig: LiR and ELR packet header size} 
\includegraphics[width=0.48\linewidth]{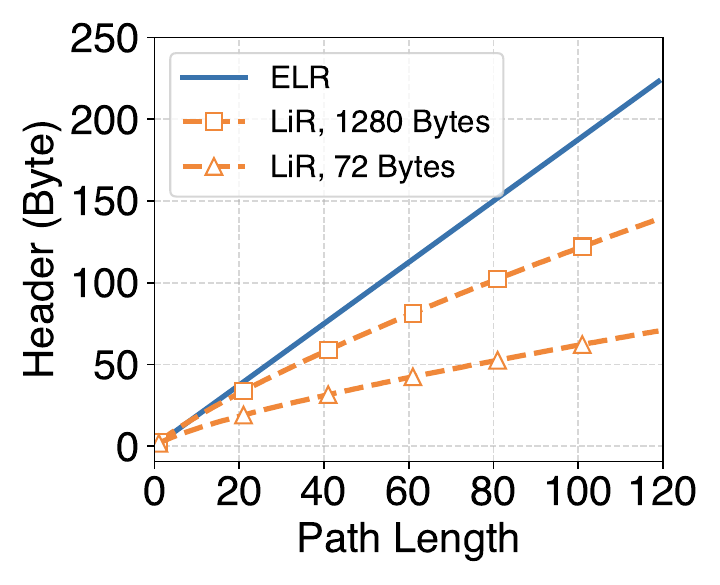}}
\subfigure[Forwarding overhead]{\label{fig: LiR and ELR forwarding overhead}
\includegraphics[width=0.48\linewidth]{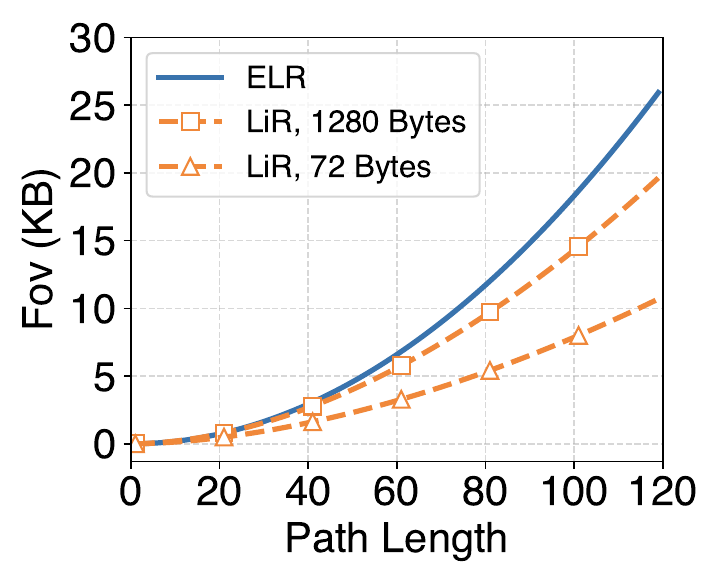}}
\caption{Numerical results of LiR and ELR}
\label{fig: Numerical Results of LiR and ELR}
\end{figure}
We evaluate these two methods under the expected Starlink constellation in March 2027, which consists of 4408 satellites \cite{2077STARLINK}.
The two sub-figures in Fig.~\ref{fig: Numerical Results of LiR and ELR} show the packet header size and forwarding overhead.
In each sub-figure, the horizontal axis represents the number of hops for the end-to-end path in the LEO constellation.
The results demonstrate that LiR is efficient in recording the end-to-end path in packets.

\section{Segment Encoding Design}
\label{Section: Encoding Scheme Optimization}
This section introduces the optimal design for the segment encoding policy under the LiR architecture.
Let us start with an overview on the major rationale. 

\subsection{Major Rationale of Segment Encoding}
\label{Subsection: Major Rationale}
The major rationale behind segment encoding is to reduce the forwarding overhead by slightly sacrificing the forwarding speed.
Consider a source-destination pair with $N$ hops in a satellite network.
The source satellite and the intermediate satellites could encode part of the ISL identifiers along the path to the destination.
In this case, the encoding satellites do not need to specify all the $N$ identifiers at a time, which reduces the forwarding overhead.
Nevertheless, the intermediate encoding satellites have to recalculate the route or look up the routing table before modifying the BF of the packet.
This will potentially increase the control overhead (in terms of the processing delay), thus sacrifices the forwarding speed.
Overall, segment encoding design is a balance between forwarding overhead and encoding delay.

Fig.~\ref{fig: Re-EncodeExample} and Fig.~\ref{fig: segment-encoding-vis} plot a four-hop data delivery under source encoding and segment encoding schemes, respectively.
\begin{itemize}
\item \textit{Source Encoding:}
In Fig.~\ref{fig: Re-EncodeExample}, the source satellite $\textit{S}_{1,1}$ encodes four ISL identifiers \{0x01, 0x05, 0x10, 0x13\} into the BF of the packet.
The intermediate satellites $\{\textit{S}_{1,2},\textit{S}_{1,3},\textit{S}_{2,3}\}$ only need to check their outgoing ISLs to determine the correct forwarding directions.
Despite the fast forwarding speed, this case may incur false positives (as shown by the red arrows). 

\item \textit{Segment Encoding:}
 In Fig.~\ref{fig: segment-encoding-vis}, source satellite $\textit{S}_{1,1}$ encodes two ISL identifiers \{0x01, 0x05\}, guiding the packets towards satellite $\textit{S}_{1,3}$. 
 The intermediate satellite $\textit{S}_{1,3}$ then empties the BF and encodes the remaining ISL identifiers \{0x10, 0x13\} into it, which incurs extra processing delay.
 In this case, the number of ISL identifiers inserted by satellites $\{\textit{S}_{1,1},\textit{S}_{1,3}\}$ is smaller than that in the source encoding case shown in Fig.~\ref{fig: Re-EncodeExample}.
 Hence there are less false positives.
\end{itemize}

Based on the aforementioned rationale, we will investigate the optimal segment encoding design.
To proceed, let us introduce the problem formulation in Section~\ref{Subsection: Segment Encoding Problem}.

\subsection{Segment Encoding Problem}
\label{Subsection: Segment Encoding Problem}
\subsubsection{Encoding Policy Model}
We characterize the ISL identifier encoding  based on a series of binary variable.
Specifically, an $\numElement$-hop packet delivery involves a total of $\numElement+1$ nodes, and we let $\mathcal{\numElement}\triangleq\{1,2,...,\numElement+1\}$ denote the set of satellites.
Moreover, $n=1$ and $n=\numElement+1$ correspond to the source satellite and destination satellite, respectively.
We let $x_{n}\in\{0,1\}$ denote the encoding decision of the $n$-th satellite.
\begin{itemize}
\item The case of $x_{n}=1$ represents that the $n$-th satellite will empty the existing BF of the received packet, and encode subsequent ISL identifiers into the BF.
This will incur extra processing delay.

\item The case of $x_{n}=0$ represents that the $n$-th satellite will directly forward the received packets (without modifying the BF).
This case will not incur extra processing delay.
\end{itemize}

Therefore, we characterize a wide range of encoding policy for the $\numElement$-hop packet delivery based on the following $(\numElement+1)$-dimensional binary vector
\begin{equation}
\begin{aligned}
    \bm{x}=
    \big(
    x_{n}\in\{0,1\}:x_1 = x_{N+1} = 1 :\forall n\in\mathcal{\numElement}
    \big).
\end{aligned}
\end{equation}
For presentation convenience, we say that the $n$-th satellite is an \textit{encoding node} if and only if $x_{n}=1$.
As one can imagine, we have $x_{1}=1$ by default, otherwise the BF is empty at the source satellite and the packet forwarding never starts.
Furthermore, $x_{\numElement+1}$ has no effect on the entire forwarding process, and we let $x_{\numElement+1}=1$ for notation simplicity.
Based on the above discussion, the encoding policy $\bm{x}$ for the $\numElement$-hop packet delivery should be chosen from the set $\mathcal{X}$, i.e.,
\begin{equation}
\begin{aligned}
    \mathcal{X}
    \triangleq
    \big( \bm{x}\in\{0,1\}^{\numElement+1}:x_{1}=x_{\numElement+1}=1 \big).
\end{aligned}
\end{equation}
Now we use a toy example to elaborate the encoding policy.

\begin{example}\label{Example: N=9}
Consider a nine-hop packet forwarding, i.e., $\numElement=9$.
The encoding policy $\bm{x}=(1,0,0,0,0,0,1,0,0,1)$ corresponds to a total of two forwarding segments.
\begin{itemize}
\item The source satellite encodes six ISL identifiers towards the destination satellite via the in-packet BF.

\item The seventh satellite will empty the BF and further encodes three ISL identifiers via the in-packet BF.
\end{itemize}
\end{example}

The above example shows that $\sum_{n=1}^{\numElement}x_{n}$ represents the number of encoding times given the encoding policy $\bm{x}$.
To facilitate our later discussion, we define an intermediate function $r_{n}(\bm{x})$ as follows:
\begin{equation}
\begin{aligned}
r_{n}(\bm{x})
\triangleq  \arg \min\limits_{i>n}&\quad i \\
\textit{s.t.}\quad&\quad x_{i}=1.
\end{aligned}
\end{equation}
Intuitively, $r_{n}(\bm{x})$ denotes the smallest index of the encoding satellites among $\{n+1,n+2,...,\numElement+1\}$.
Table \ref{table: Example} illustrates $r_{n}(\bm{x})$ based on Example \ref{Example: N=9}.
Since $x_{1}=x_{7}=x_{10}=1$ in Example \ref{Example: N=9}, we have $r_{n}(\bm{x})=7$ for any $n\in\{1,2,...,6\}$, and $r_{n}(\bm{x})=10$ for any $n\in\{7,8,...,10\}$.

\begin{table}[H]
\renewcommand{\arraystretch}{1.1}
    \setlength{\abovecaptionskip}{3pt}
    \setlength{\belowcaptionskip}{0pt}
\caption{An illustration based on Example \ref{Example: N=9}}
\label{table: Example}
\centering
\begin{tabular}{|c|c|c|c|c|c|c|c|c|c|c|}
    \hline
    Node $n$	& 1		& 2 	& 3		& 4 	& 5 	& 6 	& 7 	& 8		& 9 & 10 \\
    \hline\hline
    $\bm{x}$ 		& 1 	& 0 	& 0 	& 0		& 0		& 0		& 1		& 0		& 0 & 1 \\
    \hline
    $r_{n}(\bm{x})$	& 7		& 7		& 7		& 7		& 7		& 7		& 10	& 10		& 10 & 10 \\
    \hline
\end{tabular}
\end{table}

\subsubsection{Encoding Performance Formulation}
Given an encoding policy $\bm{x}\in\mathcal{X}$, if $x_{n}=1$, then there are a total of $r_{n}(\bm{x})-n$ ISL identifiers encoded by the $n$-th satellite.
The corresponding forwarding overhead is $f(r_{n}(\bm{x})-n)$.
Furthermore, we let $\tau$ denote the processing delay incurred at the $n$-th satellite, which depends on the route recalculation and BF updating.
We define the overhead (in delay) incurred at the $n$-th satellite as
\begin{equation}\label{Equ: delay_n}
\begin{aligned}
\left[ \frac{f(r_{n}(\bm{x})-n)}{B}+\tau \right]x_{n},
\end{aligned}
\end{equation}
where $B$ represents the bandwidth of the ISL in the satellite network.
In (\ref{Equ: delay_n}), $\tau$ represents the processing delay caused to this packet under the encoding decision $x_{n}=1$.
Moreover, ${f(r_{n}(\bm{x})-n)}/{B}$ represents the potential queuing delay caused (by the incorrect forwarding) to other packets under the encoding decision $x_{n}=1$.
Such a queuing delay may not be perceived by other packets, but incurs extra load to the satellite network.
Based on the above discussion, the total overhead (measured in delay) under the encoding policy $\bm{x}$ is
\begin{equation}\label{Equ: delay_N}
\begin{aligned}
    \sum_{n=1}^{\numElement}\left[ \frac{f(r_{n}(\bm{x})-n)}{B}+\tau \right]x_{n}.
\end{aligned}
\end{equation}
For presentation convenience, we refer to (\ref{Equ: delay_N}) as \textit{``temporal overhead''} in this paper.
Ideally, we want to design encoding policy $\bm{x}\in\mathcal{X}$ to reduce (\ref{Equ: delay_N}), leading to Problem  \ref{Problem: original}.

\begin{tcolorbox}
\begin{problem}\label{Problem: original}
Given an $\numElement$-hop packet forwarding, the optimal encoding policy $\bm{x^\star}$ is 
\begin{equation}
\begin{aligned}
    \bm{x^\star}=\arg
    \min&\quad \sum_{n=1}^{\numElement}\left[ \frac{f(r_{n}(\bm{x})-n)}{B}+\tau \right]x_{n}\\
    \textit{var.}\quad&\quad \bm{x}\in\mathcal{X}.
\end{aligned}
\end{equation}
\end{problem}
\end{tcolorbox}

Problem  \ref{Problem: original} is a binary non-linear programming, which is NP-hard in general.
Moreover, it is not monotonic or sub-modular, thus the greedy algorithm has no performance guarantee.
To overcome the challenges, we will leverage its special structure and solve Problem~\ref{Problem: original} efficiently in Section~\ref{Subsection: Encoding Policy Design}.

\subsection{Encoding Policy Design}
\label{Subsection: Encoding Policy Design}

The objective function in Problem \ref{Problem: original} exhibits a decomposable structure.
Hence one could solve it in a recursive manner.
To proceed, let us define the sub-problems.

\subsubsection{Sub-Problem Definition}
Problem \ref{Problem: sub-problem} introduces the definition of the type-$i$ sub-problem for Problem \ref{Problem: original}.

\begin{tcolorbox}
\begin{problem}[Type-$i$ Sub-Problem]\label{Problem: sub-problem}
For any $i\in\{1,2,3,...,N\}$, the type-$i$ sub-problem is as follows:
\begin{subequations}\label{Equ: H(i)}
\begin{align}
    H(i) = \min&\quad \sum_{n=1}^{i} \left[\frac{f(r_{n}(\bm{x}))-n}{B} + \tau \right] x_{n} \\
    \textit{s.t. }&\quad x_{i+1}=1 \label{Equ: H(i)_constraint} \\
    \textit{var.}&\quad \{ x_1,x_2,...,x_{i}\} \in \{0,1\}^{i}\label{Equ: H(i)_variable}.    
\end{align}
\end{subequations}
\end{problem}
\end{tcolorbox}

We let $H(i)$ denote the optimal value of the type-$i$ sub-problem.
To understand the major rationale of the type-$i$ sub-problem, let us make the following two-fold elaboration:
\begin{itemize}
\item First, (\ref{Equ: H(i)_constraint}) implies that the $(i+1)$-th satellite is presumed to be an encoding satellite in the type-$i$ sub-problem.
That is, $H(i)$ represents the minimal temporal overhead of forwarding a packet from the source satellite to the $(i+1)$-th satellite.
This way, the type-$i$ sub-problem is only related to $\{x_{1},x_{2},...,x_{i}\}$, i.e., (\ref{Equ: H(i)_variable}).

\item Second, the type-$N$ sub-problem is mathematically equivalent to Problem \ref{Problem: original}.
Hence the optimal encoding policy $\bm{x^\star}$ and the minimal temporal overhead $H(N)$ satisfy
\begin{equation}
    \begin{aligned}
        H(N) = \sum_{n=1}^{N}\left[{\frac{f(r_{n}(\bm{x^\star})-n)}{B} + \tau}\right].
    \end{aligned}
\end{equation}
\end{itemize}
The two aspects above allow us to solve Problem \ref{Problem: original} by calculating $H(i)$ for each sub-problem based on the recursive relation (to be introduced in the following).

\subsubsection{Recursive Relation}
The sub-problems exhibit a recursive relation, and the minimal temporal costs $\{H(i):1\le i\le N\}$ could be calculated efficiently.
Lemma \ref{lemma: transition} presents the recursive relation, which will be used in Algorithm~\ref{Algorithm: main}.
\begin{lemma} \label{lemma: transition}
Suppose that $H(0)=0$, then the minimal temporal cost $H(i)$ of the type-$i$ sub-problem satisfies
\begin{equation}
\label{formula: stateTransitionEquation}
\begin{aligned}
    H(i)&=\min_{0 \leq q < i} \Bigg\{ H(q) + \left[\frac{f(i-q)}{B} + \tau\right]\Bigg\}.
\end{aligned}
\end{equation}
\end{lemma}


\begin{algorithm}[t]
\caption{ISL Encoding Policy}
    \label{Algorithm: main}
\LinesNumbered 
\KwIn{Number of ISL identifiers $\numElement$}
\KwOut{The optimal encoding policy $\bm{x}^{\star}$} 
    \textbf{Initial} $H(0)=0$, $P(0)=0$, $\bm{x^\star}=(\bm{0}_{N},1)$ \label{Line: initial}\\
    \For{$i=1$ \textbf{to} $\numElement$}  { \label{Line: for_start}
    \For{$j=0$ \textbf{to} $i-1$}{
            Compute                 $\Psi(j) 
            = 
            H(j) + \left[ {\frac{f(i-j)}{B}} + \tau\right].$

        }
        \textbf{Find} $j^{\star} = \arg \min\limits_{0\leq j<i}{\Psi(j)}$ \\
        \textbf{Set} $H(i) = \Psi(j)$ and $P(i) = j^{\star}$ \label{Line: for_end} \\
} 
    \textbf{Set} $n = N$     \label{Line: Policy_start} \\
    \Repeat{$n=0$}{ \label{Line: Policy_repeat}
        Set $n=P(n)$ and $x_{n}^\star=1$    
    }\label{Line: Policy_end} 
\end{algorithm}

\subsubsection{Algorithm Design}
Based on the recursive relation in Lemma \ref{lemma: transition}, we propose an algorithm to calculate the minimal temporal costs $\{H(i):\forall 1\le i\le N\}$ and the optimal encoding policy $\bm{x^{\star}}$.
Algorithm \ref{Algorithm: main} summarizes the main procedure.

\begin{itemize}
\item {Line \ref{Line: initial}} initializes the temporal cost $H(0)=0$ and the auxiliary parameter $P(0)=0$.

\item {Lines \ref{Line: for_start}-\ref{Line: for_end}} calculate the minimal temporal cost $H(i)$ and the auxiliary vector $P(i)$ for any $i\in\{1,2,...,N\}$ based on the recursive relation.
Specifically, $P(i)$ represents the last encoding node for the type-$i$ sub-problem.

\item {Lines \ref{Line: Policy_start}-\ref{Line: Policy_end}} generate the optimal encoding policy $\bm{x^\star}$ based on the auxiliary vector $\{P(i):\forall 1\le i\le N\}$. 
Specifically, Line \ref{Line: Policy_start} means that we focus on the type-$N$ sub-problem, i.e., the original problem.
We then construct the optimal encoding policy $\bm{x^\star}$ based on the auxiliary vector in Lines \ref{Line: Policy_repeat}-\ref{Line: Policy_end}.    
\end{itemize}

\subsubsection{Algorithm Implementation}
\label{Section: Algorithm Implementation}

After introducing the algorithm design, we now introduce how to implement it in the routing and forwarding process.
For the source node, it determines the path to the destination and encodes the first segment of ISL identifiers according to the results calculated by Algorithm~\ref{Algorithm: main}.
For intermediate nodes, there are two cases for them to handle the received LiR packets.

\begin{itemize}
\item If a satellite finds itself not the destination and its outgoing ISL recorded in the BF, then it is a direct-forwarding satellite.
In this case, this satellite will forward the packets according to the in-packet BF.

\item If a satellite finds itself not the destination and none of its outgoing ISLs are recorded in the BF, then this satellite is a re-encoding satellite.
In this case, this satellite will lookup its routing table to find the path to the destination.
Once the path is identified, the satellite calculates the segment of ISL identifiers it needs to encode based on the Algorithm~\ref{Algorithm: main}.
Note that, this calculation can be done in advance to reduce the processing time.
Finally, the satellite empties the in-packet BF and encodes these identifiers into the in-packet BF to guide packet forwarding towards the next re-encoding satellite.
\end{itemize}

So far, we have introduced the optimal design for the encoding policy under the LiR architecture.
In the following, we will introduce how the LiR architecture addresses the topology dynamics in LEO constellations.
Specifically, we will present the ISL failure and GSL handover management in Section~\ref{Section: ISL Failure Management} and Section~\ref{Section: Multicast Configuration}.

\begin{figure*}
\centering
\subfigure[LSA scheme]{\label{fig: link state update failure}
\includegraphics[height=0.2\linewidth]{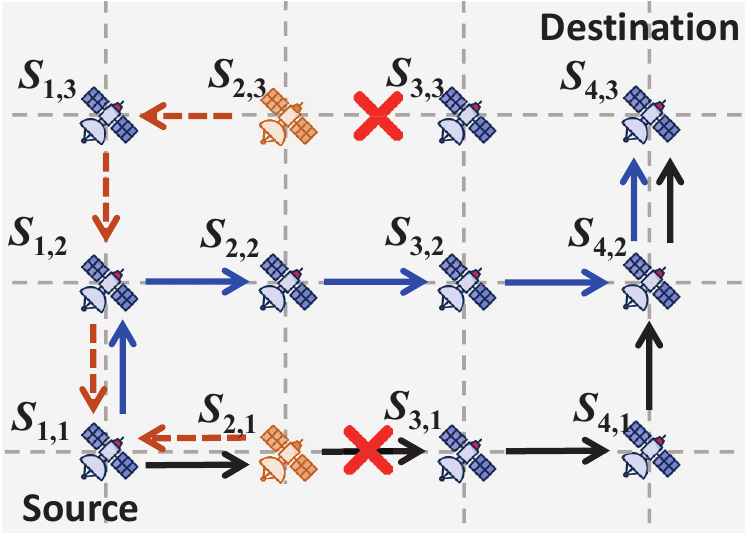}}\qquad
\subfigure[ODR Scheme]{\label{fig: rerouting link failure}
\includegraphics[height=0.2\linewidth]{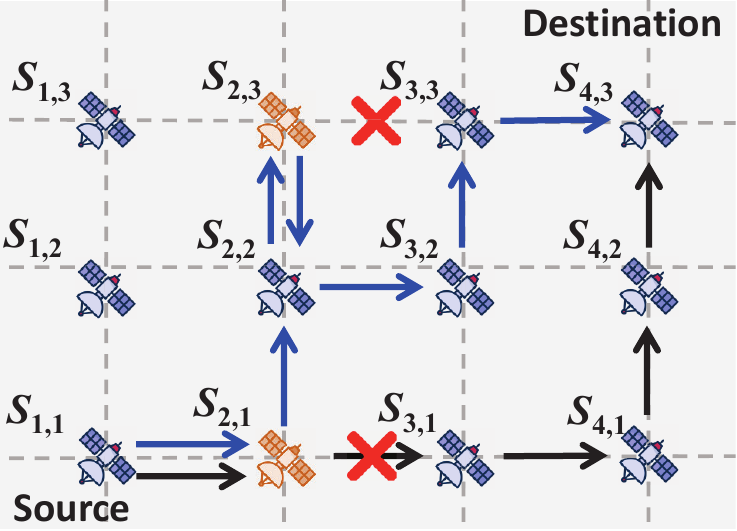}}
\subfigure[ODD Scheme]{\label{fig: virtual link failure}
\includegraphics[height=0.25\linewidth]{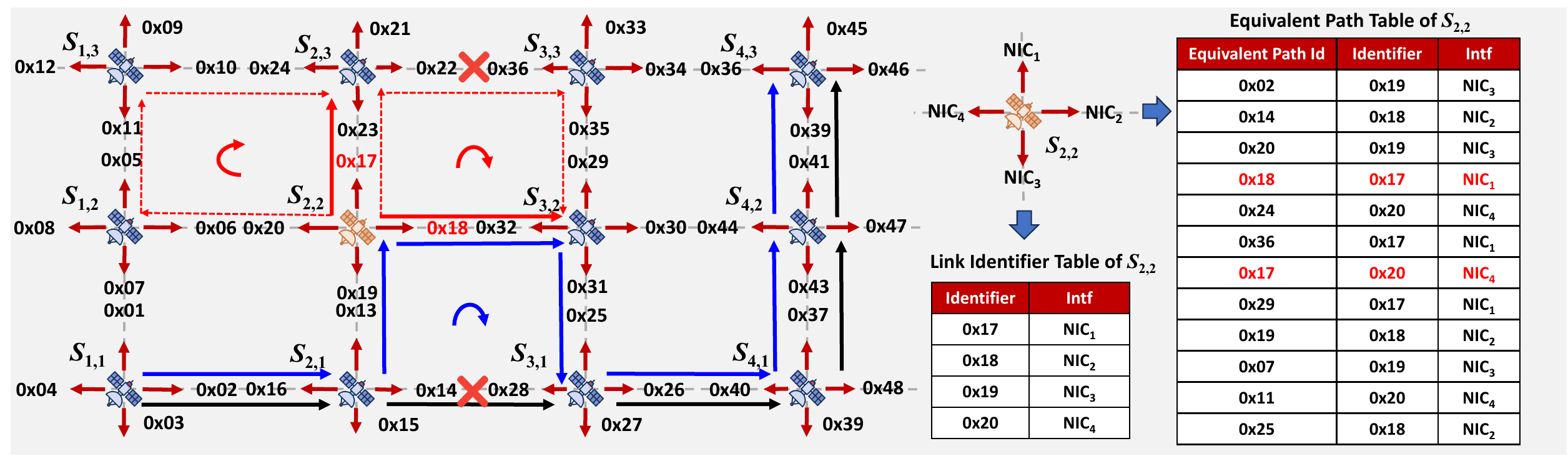}
\label{fig: ODD scheme}
}
\caption{ISL failure management. In each sub-figure, the red crosses represent ISL failures. The black arrows represent the optimal route when there is no ISL failure. The blue arrows represent the optimal route when two ISL failures occur.}
\end{figure*}

\section{ISL Failure Management of LiR}
\label{Section: ISL Failure Management}

Due to the relative movement between neighbor satellites, the ISLs are usually intermittent.
A proper ISL failure management scheme should help improve the reliability of packet forwarding.
Under LiR architecture, we consider three ISL failure management schemes.

\begin{itemize}
\item Link State Announcement (LSA) refers to the traditional link state dissemination adopted in many other routing architectures (e.g., IP-based OSPF).
Under the LSA scheme, each node periodically detect its link states, and disseminate the LSA packets in a flooding manner.

\item On-Demand Rerouting (ODR) relies on the forwarding satellites to handle the ISL failures.
Specifically, each satellite monitors the status of its ISLs in real time. 
When an intermediate satellite receives a packet that is about to be sent to a failed ISL according to the in-packet BF.
ODR recalculates the route to the destination
without this failed ISL and updates the in-packet BF

\item On-Demand Detouring (ODD) relies on the forwarding satellites to manage the ISL failures.
Each satellite monitors the status of its ISLs in real time and take actions when it receives a packet that is about to be sent to a failed ISL according to the in-packet BF.
Different from ODR scheme, ODD activates a predetermined equivalent path to bypass this failed ISL.
\end{itemize}

Note that the classic LSA scheme requires to detect the ISL state changes and disseminating the changes throughout the satellite network.
In contrast, the ODR and ODD schemes do not need to disseminate link state update packets.
Instead, they rely on the forwarding satellites to cope with their detected ISL failures via rerouting and detouring.
Next, we first provide an overview of the classic LSA scheme in Section~\ref{Subsection: LSA Scheme}.
We then elaborate how the ODR and ODD schemes work under the link-identified framework in Section~\ref{Subsection: ODR Scheme} and Section~\ref{Subsection: ODD Scheme}, respectively.
We will evaluate the performance of the three schemes in Section~\ref{Subsection: Impact of Link Failure Management}.


\subsection{LSA Scheme}
\label{Subsection: LSA Scheme}
The classic LSA scheme has been widely adopted in link-state routing protocols (e.g., OSPF) to cope with the link-state changes.
Under LSA scheme, each satellite will generate hello packets according to a fixed time interval (e.g., 5 seconds for OSPF), which will be delivered to its neighbors. 
In this case, a satellite could anticipate the link state based on whether it receives the expected hello packets from the its neighbors.

We take Fig.~\ref{fig: link state update failure} as the example to illustrate how LSA scheme works.
Specifically, we consider a source-destination pair in the satellite network, where source $\textit{S}_{1,1}$ will send packets to destination $\textit{S}_{4,3}$.
Before the data delivery process, two ISLs fail (i.e., the two red crosses) and are detected by $\textit{S}_{2,1}$ and $\textit{S}_{2,3}$.
Accordingly, these two satellites then disseminate the detected link-state changes throughout the satellite network.
The blue dash arrows represent that the source satellite $\textit{S}_{1,1}$ has perceived the two ISL failures.
Accordingly, the source satellite $\textit{S}_{1,1}$ will deliver the packets to the destination via the blue arrows to avoid the two failed ISLs.

Although the LSA scheme enables each node in the satellite network to obtain the link-state information, the link state dissemination may not be timely.
Specifically, when the source satellite has specified the path in the BF and sent out the packets, any occasional ISL failure along the path will lead to packet drops.
This indicates that the classic LSA scheme may not effectively address occasional ISL failures under LiR.
This motivates us to leverage the forwarding satellites to cope with the occasional ISL failures during the packet delivery.




\subsection{ODR Scheme}
\label{Subsection: ODR Scheme}
Under the ODR scheme, the real-time link state information will not be disseminated within the satellite network.
Instead, each satellite will calculate the route according to the deterministic neighbor relation.
Moreover, the forwarding satellites are responsible for coping with the occasional ISL failures.
Specifically, upon receiving a link-identified packet, the forwarding satellite will check the outgoing ISL encoded in the BF of the packet.
If the corresponding ISL is in a normal state, the packet is directly forwarded.
Otherwise, if the corresponding ISL is in a failure state, the forwarding satellite recalculates the route to the destination and updates the BF.

We take Fig.~\ref{fig: rerouting link failure} as the example to illustrate how the ODR scheme works.
Similarly, we consider the packet delivery from source satellite $\textit{S}_{1,1}$ to destination satellite $\textit{S}_{4,3}$.
Recall that the ODR scheme will not disseminate the link-state changes.
Hence the source satellite cannot perceive the two failed ISLs in this case.
The black arrows represent the route specified by the source satellite via the in-packet BF.
When the packet reaches the intermediate satellite $\textit{S}_{2,1}$, it recalculates the route (i.e., $\textit{S}_{2,1}\rightarrow\textit{S}_{2,2}\rightarrow\textit{S}_{2,3}\rightarrow\textit{S}_{3,3}\rightarrow\textit{S}_{4,3}$) and updates the in-packet BF.
When the packet reaches the satellite $\textit{S}_{2,3}$, it detects that the ISL between $\textit{S}_{2,3}$ and $\textit{S}_{3,3}$ has failed.
It then performs the rerouting again, resulting in a new route (i.e., $\textit{S}_{2,3}\rightarrow\textit{S}_{2,2}\rightarrow\textit{S}_{3,2}\rightarrow\textit{S}_{3,3}\rightarrow\textit{S}_{4,3}$).
Finally, the packet reaches the destination satellite $\textit{S}_{4,3}$.

\subsection{ODD Scheme}
\label{Subsection: ODD Scheme}
Similar to the ODR scheme, the ODD scheme will not disseminate the link state information within the satellite network.
Instead, the forwarding satellites will cope with the occasional ISL failures by configuring an equivalent path via the BF in an on-demand manner.
An equivalent path in LiR-ODD refers to the shortest path in terms of hops to bypass a failed link identifier and reach the other side.
Given the grid-like topology, the equivalent path may follow a clockwise or counterclockwise direction.
As shown in Fig.~\ref {fig: ODD scheme}, the clockwise equivalent path of the ISL 0x14 is the path $\textit{S}_{2,1}\rightarrow\textit{S}_{2,2}\rightarrow\textit{S}_{3,2}\rightarrow\textit{S}_{3,1}$.
When satellite $\textit{S}_{2,1}$ receives the packet and finds that the outgoing ISL 0x14 has failed, it configures the aforementioned equivalent path under the LiR architecture.
To achieve this goal, we need to address three questions: 
a) how to calculate equivalent paths;
b) how to represent equivalent paths in the packets; 
c) how to forward the packet along an equivalent path.
We address the three questions as follows:
\begin{itemize}
\item An equivalent path refers to the shortest path in terms of hops to bypass a failed link identifier.
Given the grid-like topology, the equivalent path may follow a clockwise or counterclockwise direction.
Under LiR architecture, the equivalent path is represented by the link identifiers it contains.
As shown in Fig.~\ref{fig: ODD scheme}, we plot two clockwise equivalent paths of two link identifiers (i.e., one red inter-orbit link identifier and one red intra-orbit link identifier).

\item We represent an equivalent path with a single identifier (i.e., the link identifier itself).
Moreover, to differentiate it from normal identifiers and maintain a low false positive rate of the normal BF, we store this identifier in a different BF named equivalent-path BF.

\item Although the activated equivalent paths are encoded into the equivalent-path BF, intermediate satellites cannot determine which interface to forward the packets.
To address this, each satellite must maintain an equivalent path forwarding table, which maps the equivalent path identifier to the outgoing interface of this satellite.
As shown in Fig.~\ref{fig: ODD scheme}, the clockwise equivalent path of link identifier 0x14 comprises three link identifiers \{0x13, 0x18, 0x31\}, among these link identifiers the link identifier 0x18 belongs to the satellite $\textit{S}_{2,2}$.
Consequently, $\textit{S}_{2,2}$'s equivalent path table should include an entry that maps the equivalent path identifier 0x14 to the outbound link identifier 0x18, which corresponds to $\text{NIC}_{2}$.
\end{itemize}
This way, a packet with equivalent BF could avoid the failed ISL by detouring a little.
We take Fig.~\ref{fig: ODD scheme} as an example to illustrate how the ODD scheme works.
Upon receiving a packet from $\textit{S}_{1,1}$, satellite $\textit{S}_{2,1}$ will first lookup the identifiers in regular forwarding table and check the normal BF to obtain the forwarding direction.
However, the designated link identifier 0x14 is failed.
Subsequently, the satellite $\textit{S}_{2,1}$ takes the responsibility of configuring the equivalent path for the failed ISL identifier 0x14 by encoding the 0x14 into the equivalent-path BF.
Specifically, the equivalent path of the failed ISL consists of three ISL identifiers (i.e., 0x13, 0x18, 0x31).
When $\textit{S}_{2,2}$ receives the packet, it lookup its equivalent path table and find 0x14 is encoded into the equivalent BF.
Consequently, the packet should be forwarded from $\text{NIC}_{2}$.
Eventually, when the packet reaches $\textit{S}_{3,1}$, the equivalent-path BF will be removed and the packet will be further forwarded according to the original route.



\section{GSL Handover Management of LiR}
\label{Section: Multicast Configuration}
In LEO satellite networks, frequent GSL handover will increase the topology dynamics.
Hence GSL handover management plays a crucial role on LEO satellite network routing.
In Section~\ref{Subsection: Advantages in Satellite-Ground Handover}, we introduce how we address the impact of the GSL handover under LiR by leveraging multicast forwarding.
In Section~\ref{Subsection: multicast mechanism}, we then present two types of multicast mechanism under LiR architecture.





\begin{figure}
\centering
\includegraphics[width=0.95\linewidth]{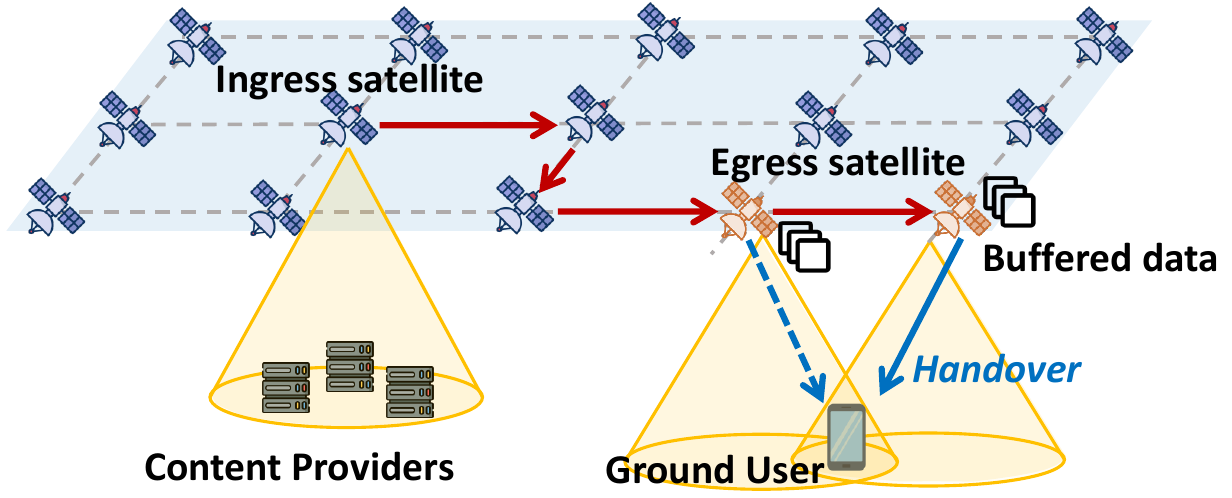}
\caption{Illustration of multicasting in satellite handover}
\label{fig: multicast advantage}
\end{figure}

\subsection{GSL Handover via LiR Multicast}
\label{Subsection: Advantages in Satellite-Ground Handover}
Now we elaborate how to leverage multicast forwarding to achieve seamless GSL handover.
Recall that the orbital period of LEO satellites is short, which leads to frequent GSL handovers. 
Frequent GSL handovers reduce topology stability, and decrease the packet delivery ratio between the ingress satellite and the egress satellite. 
As shown in Fig.~\ref{fig: multicast advantage}, several ground content providers are delivering packets to a ground user via the LEO satellite network.
If the GSL handover at the user side is not disseminated to the satellite networks in time, then the packets may eventually reach a satellite that is not connecting to the destination user, leading to packet loss.
One of the widely adopted methods for addressing frequent GSL handovers is delivering the packets to multiple satellites that can cover the destination ground user in a multicast manner~\cite{UserCentricHandover}.
The validity is two-fold:
\begin{itemize}
\item First, these satellites could buffer the received packets in case GSL handover suddenly occurs.

\item Second, an efficient multicast scheme can reduce redundant packet delivery, thus incurs little overhead to achieve seamless GSL handover.
\end{itemize}

In general, the goal of seamless GSL handover via multicasting could be achieved under the classic IP architecture.
However, achieving this requires frequent reconfiguration of multicast group addresses, as the satellites covering the ground user change frequently.
In contrast, our proposed LiR architecture naturally supports multicast transmission.
The ingress satellite could specify the multicast route to the egress satellites via the in-packet BF.
Details are given in Section~\ref{Subsection: multicast mechanism}.

\begin{figure}
\centering
\subfigure [Shortest-path-first multicast]{\label{fig: spf multicast}
\includegraphics[width=0.45\linewidth]{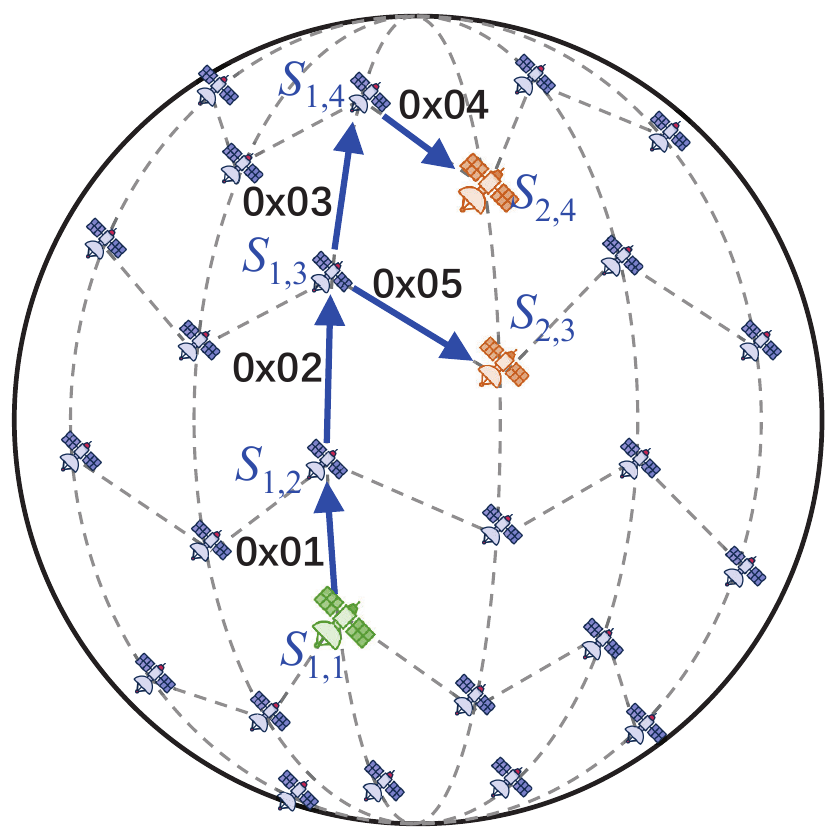}}
\subfigure[Primary-node based multicast]{\label{fig: primary multicast}
\includegraphics[width=0.45\linewidth]{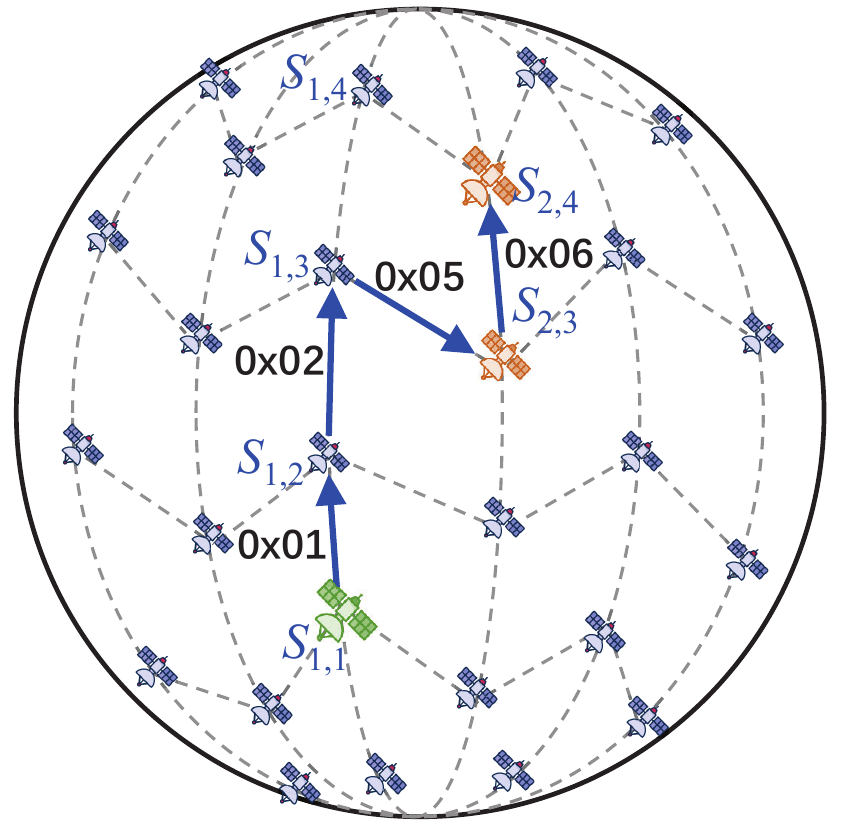}}
\caption{Different multicast mechanisms.}
\label{fig: different multicast mechanism}
\end{figure}

\subsection{Multicast under LiR}
\label{Subsection: multicast mechanism}
Different from the classic IP architecture, our proposed LiR architecture identifies each ISL in the satellite network. 
This way, the source satellite can specify the multicast tree based on the ISL identifiers. 
The intermediate satellites only need to forward the packets based on the encoded ISL identifiers.
Therefore, under LiR architecture, the transition from unicast to multicast is quite straightforward.
There is no need to develop extra protocols to configure the multicast group address (as the classic IP architecture does).
Next we will introduce different mechanisms for calculating the multicast routes.

\begin{figure*}
\centering
\subfigure[Impact of BF size $\lenBF$ on False Positive Rate (FPR) $p(M,N,K)$ under $K=5$]{\label{fig: Verification_theoretical}
\includegraphics[height=0.16\linewidth]{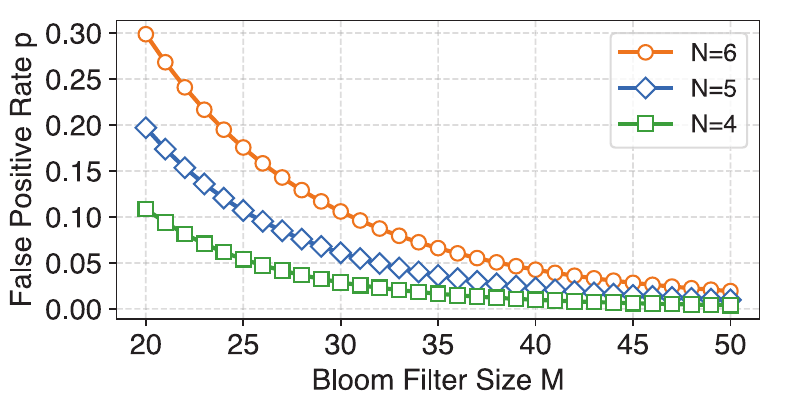}}\quad
\subfigure[Verification by randomly generating the incorrect forwarding events]{\label{fig: Verification_matlab}
\includegraphics[height=0.16\linewidth]{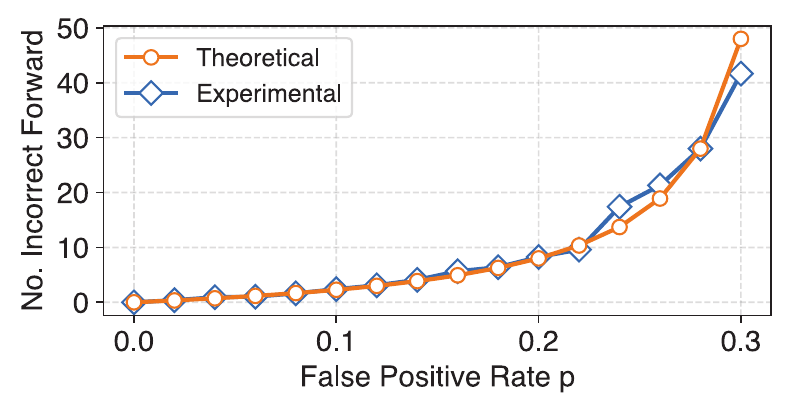}}\quad
\subfigure[Verification via the packet-level experiments on OMNeT++]{\label{fig: Verification_omnet}
\includegraphics[height=0.16\linewidth]{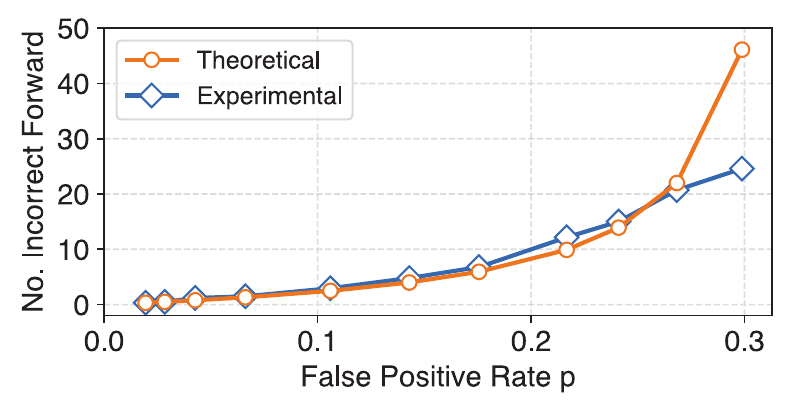}}
\caption{Verification of Theorem \ref{Theorem: incorrect}. }
\label{fig: verification}
\end{figure*}

\textbf{Shortest-Path-First (SPF) Multicast:} 
If the source satellite aims to reduce propagation delay, then it could adopt the SPF multicast mechanism, which works as follows:
\begin{itemize}
\item The source satellite computes the shortest path to each destination satellite according to Dijkstra algorithm \cite{dijkstra1959note}.
Each path is associated with a set of ISL identifiers.

\item The source satellite then takes the union of the aforementioned sets, and encodes the corresponding ISL identifiers into the in-packet BF.
\end{itemize}
As shown in Fig.~\ref{fig: spf multicast}, the source satellite $\textit{S}_{1,1}$ is going to deliver packets to satellites $\textit{S}_{2,3}$ and $\textit{S}_{2,4}$.
Under SPF multicast mechanism, the source satellite $\textit{S}_{1,1}$ calculates the shortest paths to the destination satellites (i.e. $\textit{S}_{2,3}$ and $\textit{S}_{2,4}$), and obtains two routes, i.e., \{0x01, 0x02, 0x03, 0x04\} and \{0x01, 0x02, 0x05\}.
The union of the two sets is \{0x01, 0x02, 0x03, 0x04, 0x05\}, which will be encoded into the BF.

\textbf{Primary-Node-based (PNB) Multicast:}
When the destination satellites are closed to each other, the source satellite could leverage the PNB multicast mechanism to reduce redundant delivery.
It works as follows:
\begin{itemize}
\item  The source satellite selects a satellite among its destination as the primary node.
\item The source satellite calculates the shortest path between itself and the primary node as well as the shortest paths from the primary node to the other destinations. 

\item The source satellite takes the union of the ISL identifiers in aforementioned routes, and encodes them into the BF.
\end{itemize}
We take Fig.~\ref{fig: primary multicast} as the example.
The source satellite $\textit{S}_{1,1}$ selects satellite $\textit{S}_{2,3}$ as the primary node. 
The shortest path from the source satellite $\textit{S}_{1,1}$ to the primary node $\textit{S}_{2,3}$ is \{0x01, 0x02, 0x05\}.
The shortest path from the primary node to the other destination satellite $\textit{S}_{2,4}$ is \{0x06\}.
Finally, the source satellite will encode the ISL identifiers \{0x01, 0x02, 0x05, 0x06\} into the BF of the packets to be delivered.

\section{Experimental Results} 
\label{Section: Experimental Results}

We conduct extensive packet-level experiments to verify our analytical results and evaluate our proposed LiR architecture in LEO constellations.
Specifically, we implement LiR in INET 4.2.2 (i.e., an open-source library of OMNeT++). 
Our packet-level experiments are based on  Iridium constellation, which is a typical polar constellation consisting of 66 LEO satellites at an altitude of 780km.
The bandwidth of each ISL is set to 10 Mbps, and the time consumption of BF replacement is set to $\tau=10$ microseconds. 
Additionally, the effective data volume of each packet is $C=1$ KB.
Next, we will first introduce our simulation environment based on OMNeT++ in Section~\ref{Subsection: Simulation Environment}.
Following this, we will verify our analytical results in Section~\ref{Subsection: Verification}.
We then provide the experimental results of single-flow and multi-flow cases in Section~\ref{Subsection: Performance Evaluation on Single Flow} and Section~\ref{Subsection: Performance Evaluation on Multiple Flows}, respectively.
We assess the performance of SRv6 and LiR in Section{~\ref{Subsection: Performance Evaluation on SRv6 and LiR}} and evaluate different ISL failure management schemes in Section~\ref{Subsection: Impact of Link Failure Management}.
Finally, we demonstrate the performance of multicasting under LiR architecture in Section~\ref{Subsection: Multicast Transmission under LiR}.


\subsection{Simulation Environment}
\label{Subsection: Simulation Environment}
To simulate the satellite network and conduct packet-level experiments, we rely on OMNeT++ and INET to develop an experimental platform to evaluate the routing performance.
Our platform consists of three key modules, i.e., constellation module, network module and traffic generation module.

\begin{itemize}
\item \textbf{Constellation Module} initializes the satellite nodes and ISLs according to specific simulation parameters.
To simulate satellite movement, we use the OsgEarth library to obtain real-time satellite positions, allowing us to determine ISL propagation delays.
Additionally, to simulate the topology dynamics, this module will randomly generate ISL failure and recovery events according to Poisson process with specified failure rates.

\item \textbf{Network Module}
is used to simulate the specific packet-level events, including protocol stacks, queuing policy. We implement our proposed routing architecture (i.e., LiR), ISL failure management schemes and multicast schemes in this module.

\item \textbf{Traffic Generation Module} is used for network traffic simulation under different traffic patterns (i.e., single-flow traffic pattern and one-to-many traffic pattern). 
We also take into account the traffic intensity and traffic flow duration between the providers and users.
\end{itemize}


\subsection{Performance Verification}
\label{Subsection: Verification}
Recall that Theorem \ref{Theorem: incorrect} presents the analytical result of the incorrect forwarding overhead in a closed-form.
The encoding policy design in Problem \ref{Problem: original} is also based on this result.
Now we verify the correctness of Theorem \ref{Theorem: incorrect} via extensive simulations. 
Specifically, Fig.~\ref{fig: verification} plots the number of incorrect forwarding under different false positive rates. The two sub-figures correspond to two types of experiments.

In Fig.~{\ref{fig: Verification_theoretical}}, we explore the false positive rate of bloom filter (BF) under different BF sizes.
The three curves correspond to different numbers of link identifiers, respectively.
Comparing the three curves shows that encoding more link identifiers leads to a higher false positive rate.
In addition, we vary the false positive rate of BF by changing the BF length from $20$ to $50$ bits.
Note that a larger BF results in a lower false positive rate.

%

In Fig.~\ref{fig: Verification_matlab}, we evaluate the number of incorrect forwarding hops by simulating the false positive events according to the given rates $[0,0.3]$.
False positive events are simulated using random number generation on Matlab, which does not take into account specific network topology or other practical issue.
In Fig.~\ref{fig: Verification_matlab}, the orange circle curve corresponds to the theoretical result of Theorem~\ref{Theorem: incorrect}, and the blue diamond curve corresponds to the experimental result on Matlab (under 500 simulation runs). 
Note that the two curves almost overlap under different false positive rates, which implies the correctness of Theorem \ref{Theorem: incorrect}.

In Fig.~\ref{fig: Verification_omnet}, we quantify the number of incorrect forwarding hops by simulating a single-packet transmission job in Iridium constellation on OMNeT++.
We vary the false positive rate by changing the BF length $\{20,21,22,23,25,27,30,35,40,45,50\}$ (in bit).
Overall, the two curves in Fig.~\ref{fig: Verification_omnet} almost overlap when the false positive rate is smaller than 0.27.
When the false positive rate is greater than 0.27, however, the number of incorrect forwarding hops under OMNeT++ simulation is smaller than that of the theoretical result.
This can be attributed to different handling of loops between the Matlab and OMNeT implementation.

\begin{figure}
\centering
\subfigure[Numerical result on Matlab]{\label{fig: single-flow-matlab}
\includegraphics[width=0.48\linewidth]{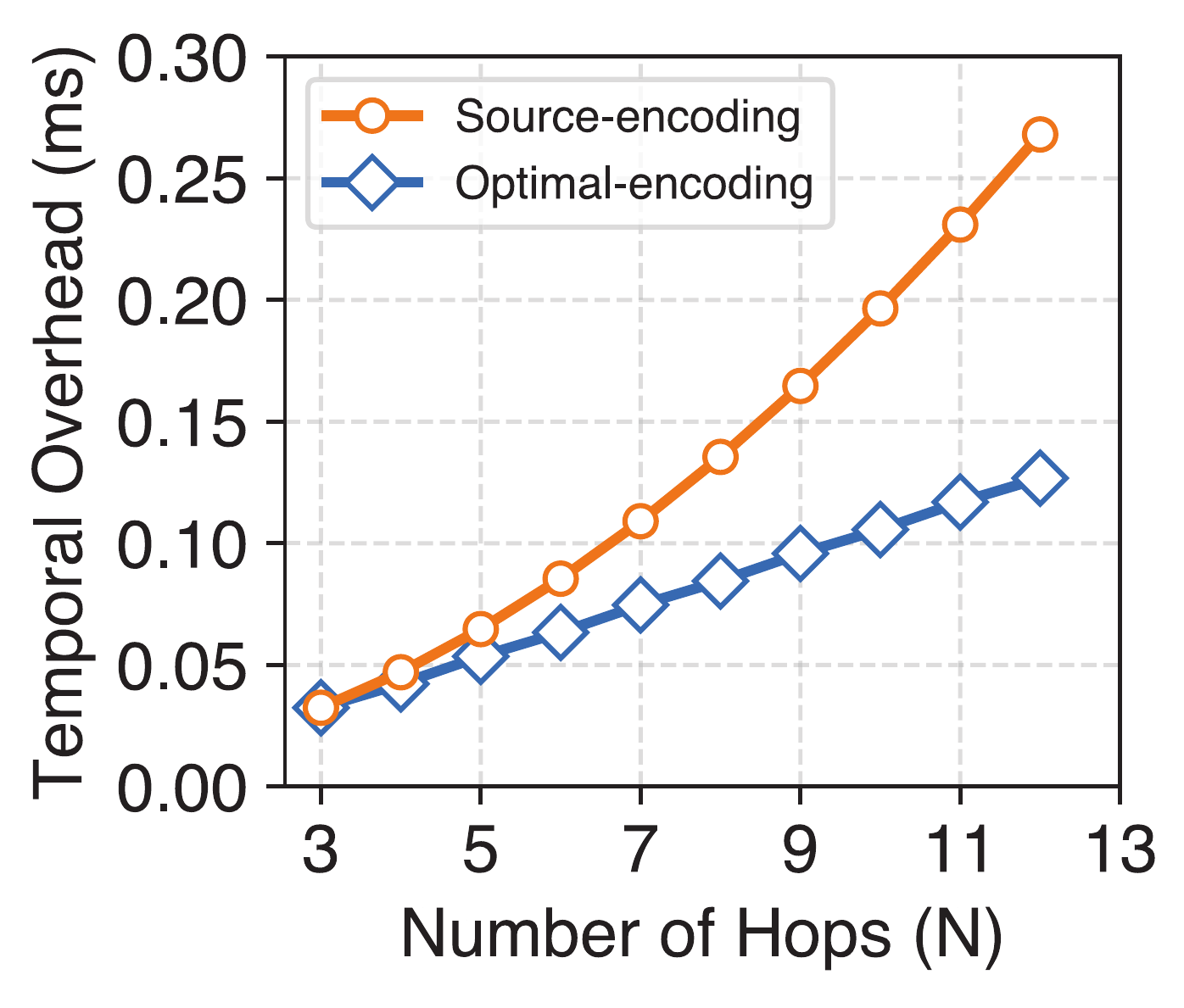}}
\subfigure[Packet-level result on OMNeT++]{\label{fig: single-flow-omnet}
\includegraphics[width=0.48\linewidth]{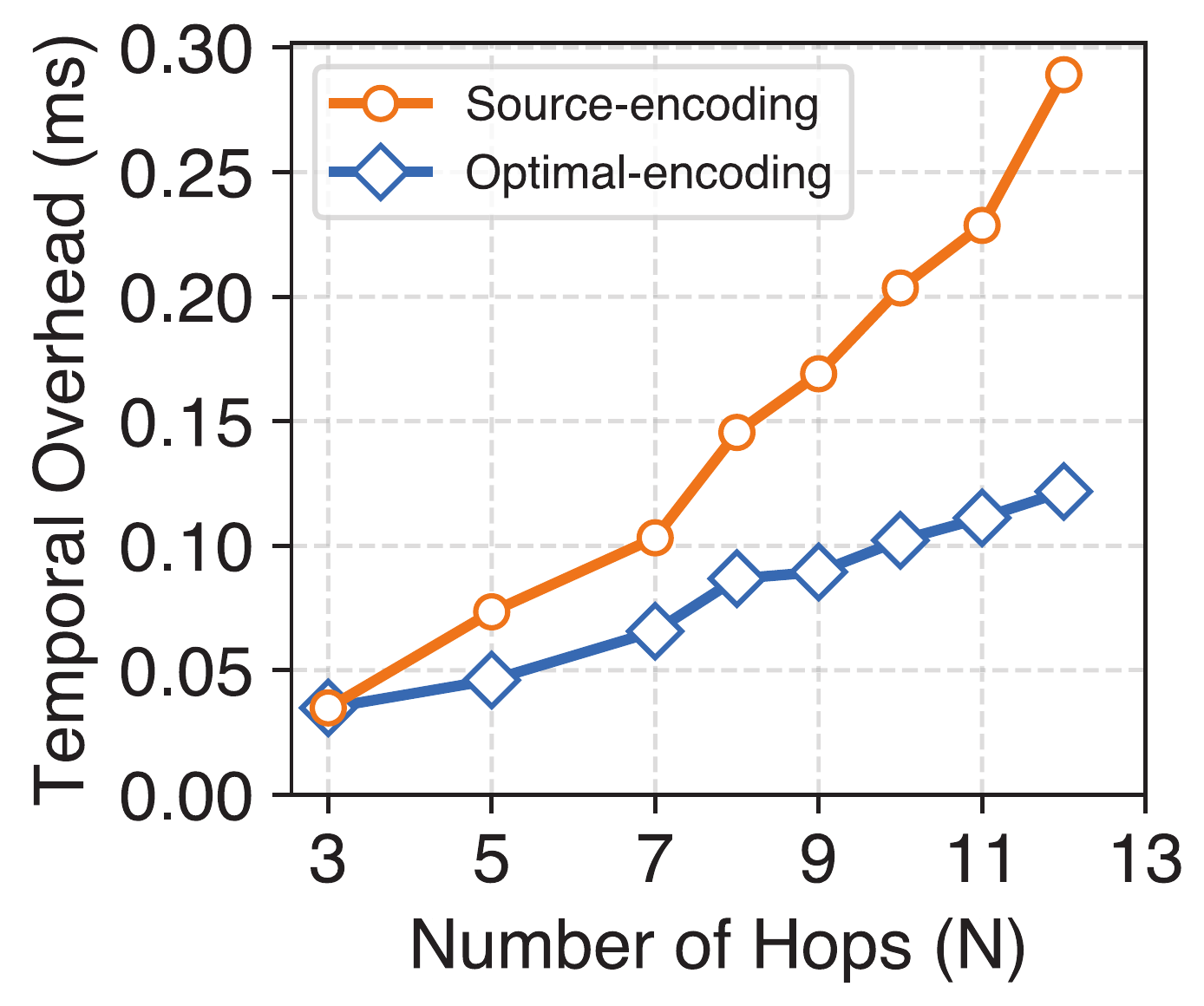}}
\caption{Performance evaluation of single-flow scenario}
\label{fig: single-flow}
\end{figure}

\subsection{Performance Evaluation on Single Flow}
\label{Subsection: Performance Evaluation on Single Flow}

We demonstrate the performance of the optimal encoding policy $\bm{x^\star}$ in a single-flow scenario.
Specifically, we vary the number of hops (i.e., $N\in\{3,4,...,12\}$) of the source-destination pair, and evaluate the temporal overhead defined in (\ref{Equ: delay_N}).
The two sub-figures in Fig.~\ref{fig: single-flow} plot the results under two types of simulations.
In each sub-figure, the horizontal axis and the vertical axis represent the number of hops and the temporal overhead (defined in (\ref{Equ: delay_N})), respectively.

Fig.~\ref{fig: single-flow-matlab} shows the numerical results generated by Matlab.
The two curves in Fig.~\ref{fig: single-flow-matlab} represent the source encoding policy and the optimal segment encoding policy $\bm{x^\star}$.
The temporal overhead gap between the two policies is increasing as the distance of the source-destination pair increases, which unveils the necessity of our proposed encoding policy design.
Furthermore, Fig.~\ref{fig: single-flow-omnet} shows the packet-level experimental results under the Iridium constellation on OMNeT++. 
Comparing Fig.~\ref{fig: single-flow-omnet} to Fig.~\ref{fig: single-flow-matlab} indicates that our theoretical results for the optimal encoding policy design are consistent in the packet-level experiments.

\subsection{Performance Evaluation on Multiple Flows}
\label{Subsection: Performance Evaluation on Multiple Flows}

\begin{figure}
\centering
\subfigure[Queuing delay]{\label{fig: average queue delay}
\includegraphics[width=0.48\linewidth]{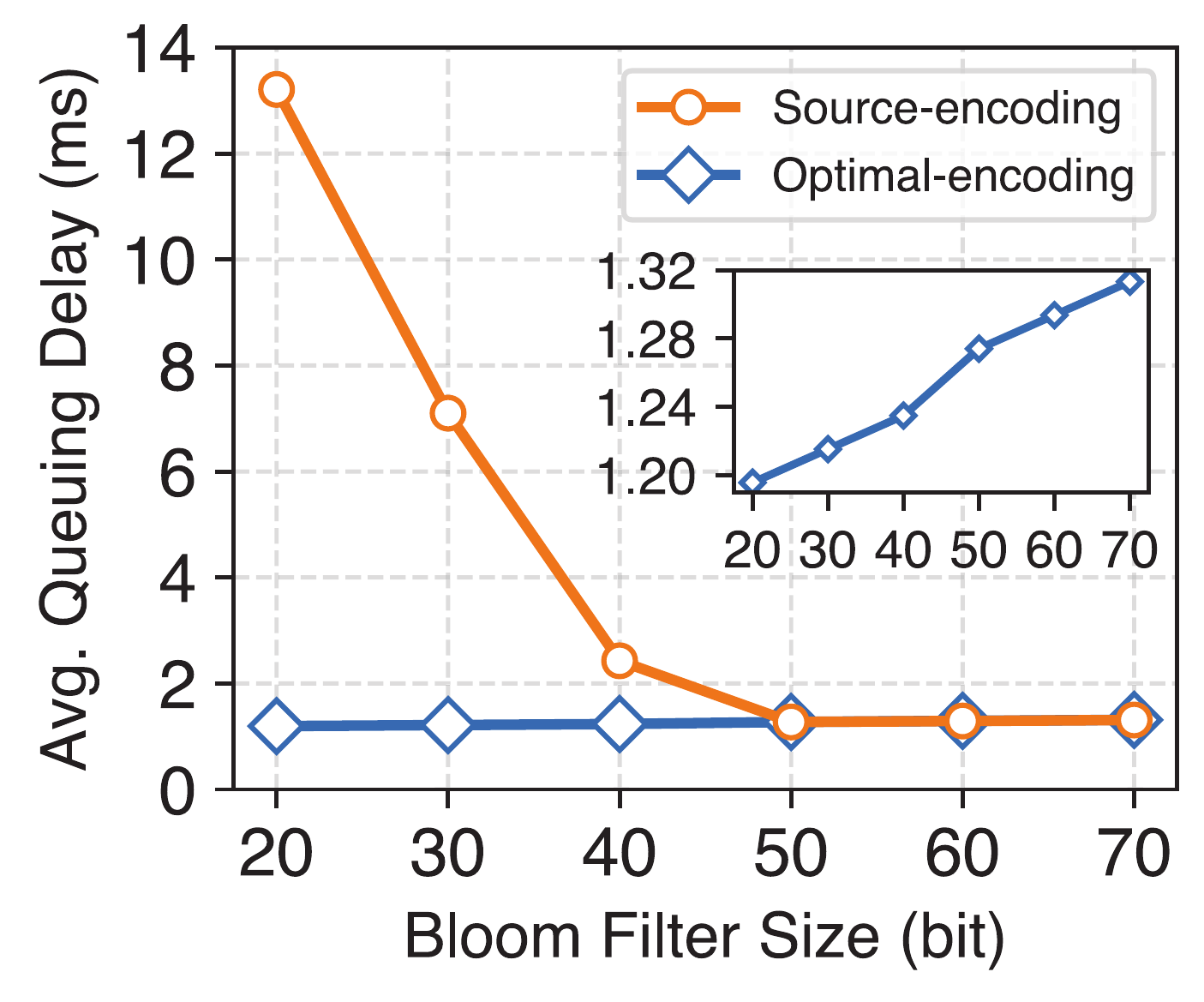}}
\subfigure[End-to-end delay]{\label{fig: average end-to-end delay}
\includegraphics[width=0.48\linewidth]{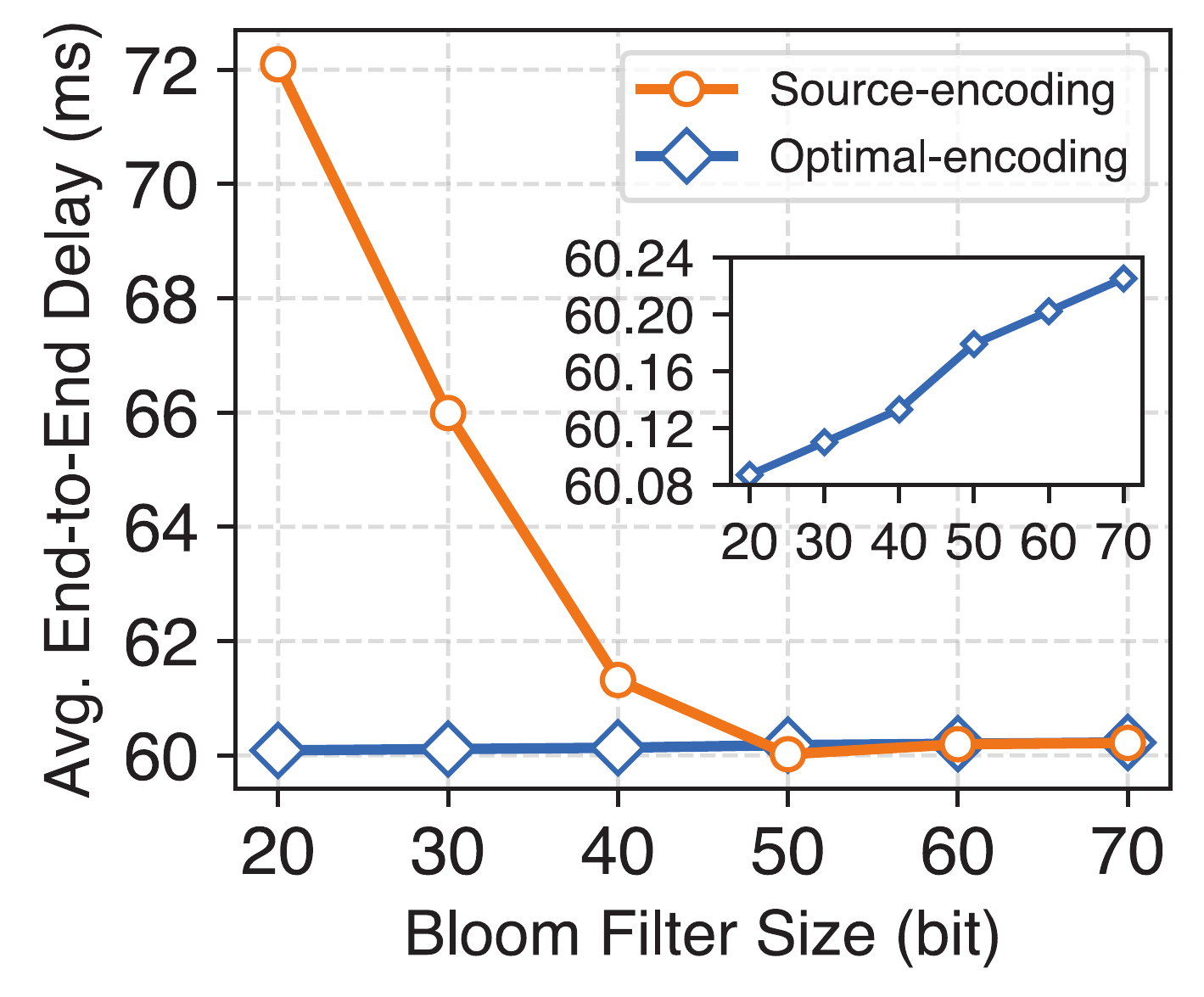}}
\caption{Performance evaluation on multiple flows}
\label{fig: multi-flow}
\end{figure}

We take into account multiple flows and evaluate the end-to-end transmission performance.
Specifically, we consider four pairs of two-way source-destination with overlapping ISLs.
The sending rate of each source satellite is 1250 packets per second. 
Note that the incorrect forwarding (caused by false positives) could potentially increase the queuing delay of other flows.
Fig.~\ref{fig: multi-flow} plots the results under different BF length $\lenBF\in\{30,40,...,70\}$ (measured in bit).
In this experiment, we do not take the BF size $\lenBF$ as a design space.

Fig.~\ref{fig: average queue delay} plots the average queuing delay across all the transmission pairs.
The two curves in Fig.~\ref{fig: average queue delay} correspond to the source encoding policy and the optimal segment encoding policy, respectively.
Note that the optimal encoding policy $\bm{x^\star}$ significantly outperform the source-encoding policy when the BF size is small.
The two curves will converge to the same value, i.e., the queuing delay without incorrect forwarding.
The blue diamond curve is slightly increasing.
This is because of the increment of traffic load due to the BF size.
Fig.~\ref{fig: average end-to-end delay} plots the end-to-end latency, which includes the queuing delay, propagation delay, transmission delay, and the ISL encoding delay.
The key insights are similar to that of Fig.~\ref{fig: average queue delay}.

\begin{figure}
\centering
\subfigure[Length of header]{\label{fig: length of header}
\includegraphics[width=0.48\linewidth]{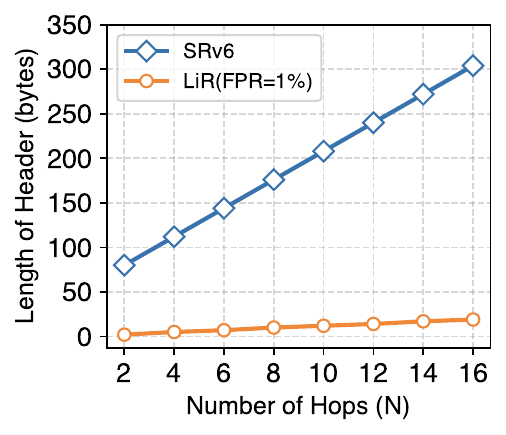}}
\subfigure[End-to-end delay]{\label{fig: transmission delay of header}
\includegraphics[width=0.48\linewidth]{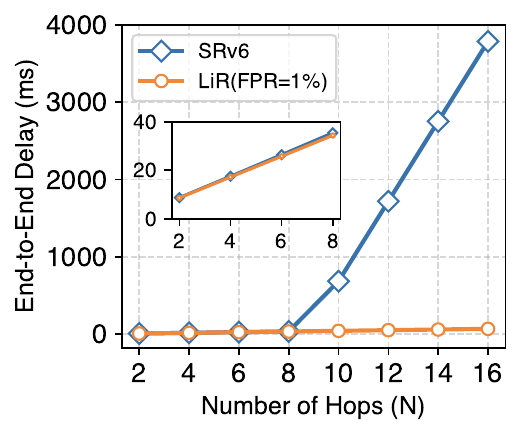}}
\caption{Performance evaluation of SRv6 and LiR}
\label{fig: SR LIR OMNeT++ Experiments}
\end{figure}

\subsection{Performance Evaluation on SRv6 and LiR}
\label{Subsection: Performance Evaluation on SRv6 and LiR}
In addition to the numerical results provided in Section~\ref{Section: Comparison Between LiR and SRv6}, we also conduct packet-level experiments to evaluate the performance of SRv6 and LiR.
In each sub-figure, the blue diamond curve represents SRv6.
The orange circle curve corresponds to LiR when the false positive rate of the in-packet BF is 1\%.
The payload of each packet is set to 1KB.
The sending rate of source satellite is 1000 packets per second.
Fig.~\ref{fig: length of header} shows the length of packet header under SRv6 and LiR.
As the number of hops increases, the length of SRv6 packet header increases faster than that of LiR header.
{Fig.~\ref{fig: transmission delay of header}} shows the end-to-end delay under SRv6 and LiR.
When the number of hops is less than 8, the end-to-end delay of SRv6 is slightly greater than that of LiR.
However, when the number of hops exceeds 8, the end-to-end delay of SRv6 significantly increases.
This is because the increase in packet header length causes congestion under SRv6.



\subsection{ISL Failure Management}
\label{Subsection: Impact of Link Failure Management}

\begin{figure}
\centering
\subfigure[End-to-End packet delay]{\label{fig: failure average end-to-end delay}
\includegraphics[width=0.95\linewidth]{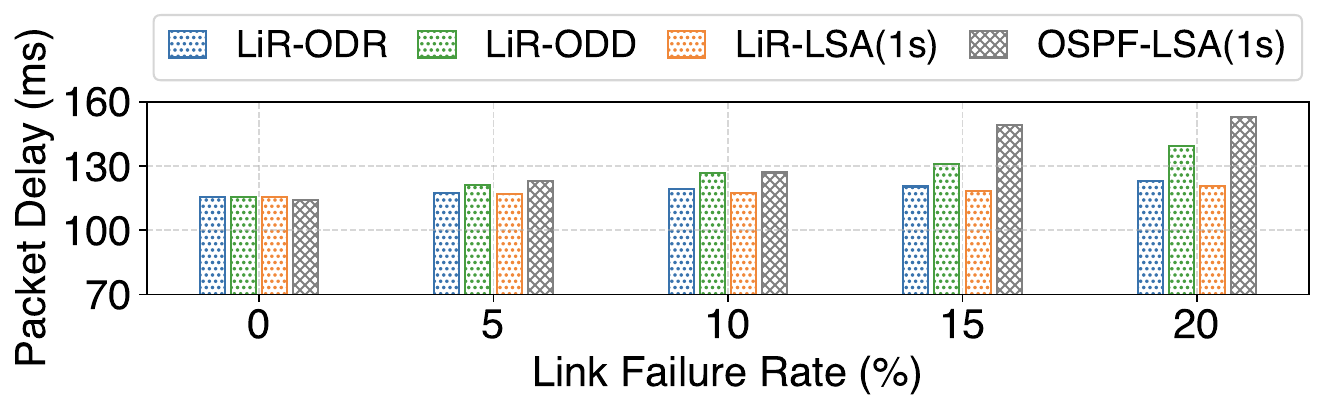}}
\subfigure[Packet delivery ratio]{\label{fig: failure successful ratio}
\includegraphics[width=0.95\linewidth]{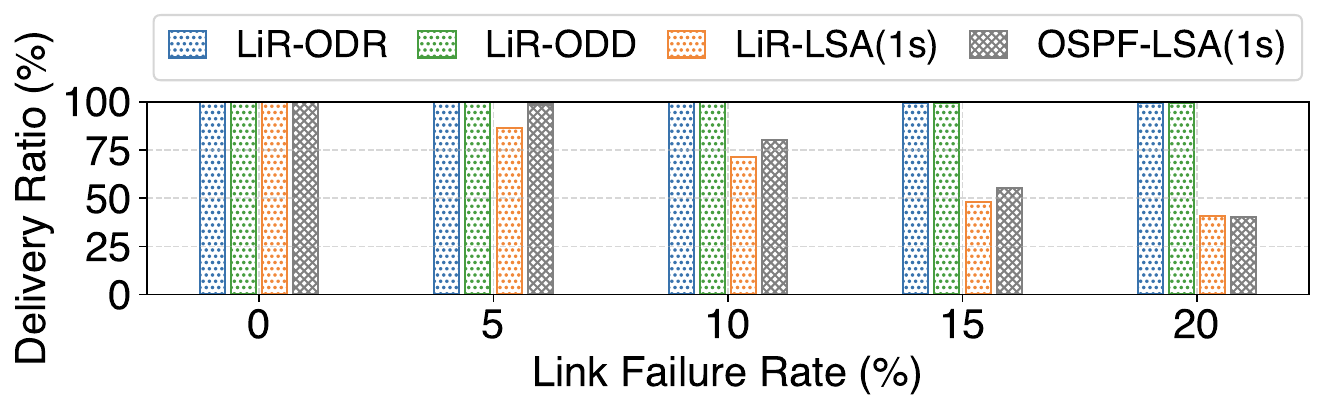}}
\caption{Performance of ISL failure management schemes}
\label{fig: link failure mechanism}
\end{figure}

We evaluate the performance of different ISL failure management schemes.
Specifically, we will evaluate the end-to-end delay and packet delivery ratio of a source-destination pair under different inter-satellite links failure rates, i.e., \{0\%, 5\%, 10\%, 15\%, 20\%\}.
The sending rate of the source satellite is 100 packets per second.
Fig.~\ref{fig: link failure mechanism} shows the results of different ISL failure management schemes.
\begin{itemize}
\item The case of LiR-ODR corresponds to the LiR architecture with the on-demand rerouting (ODR) scheme.
The blue bars in the Fig.~\ref{fig: link failure mechanism} show the results of this case.

\item The case of LiR-ODD corresponds to the LiR architecture with the on-demand detouring (ODD) scheme.
The green bars in the Fig.~\ref{fig: link failure mechanism} show the results of this case.

\item The case of LiR-LSA(1s) in Fig.~\ref{fig: link failure mechanism} corresponds to the LiR architecture with link state announcement (LSA) scheme.
The interval of sending hello packets is a second. 
The orange bars in Fig.~\ref{fig: link failure mechanism} show the results of this case.

\item The case of OSPF-LSA(1s) corresponds to the IP-based OSPF routing with link state announcement (LSA) scheme.
The hello interval is set to 1 second. 
The gray bars in Fig.~\ref{fig: link failure mechanism} show the results of this case.
\end{itemize}

Based on Fig.~\ref{fig: link failure mechanism}, we obtain three observations.

\textbf{Observation~1:}
Under the LiR architecture, the two cases LiR-ODR and LiR-ODD achieve almost the same packet delivery ratio.
However, LiR-ODR outperforms LiR-ODD in terms of the end-to-end delay.
This is because LiR-ODD relies on equivalent path to cope with occasional ISL failures, while the equivalent path may not be the delay-minimizing one.
In contrast, LiR-ODR copes with occasional ISL failures by recalculating the optimal route towards the destination.
Hence LiR-ODR will lead to a smaller end-to-end delay.


\textbf{Observation~2:} 
Given the same LSA, OSPF-LSA(1s) achieves a higher packet delivery ratio but incurs a larger end-to-end delay than LiR-LSA(1s).
Recall that OSPF-LSA(1s) makes forwarding decisions at each intermediate satellite.
In contrast, LiR-LSA(1s) allows the source satellite to specify the end-to-end path.
If LSA is not in time, the two cases will face different drawbacks:
\begin{itemize}
\item Under LiR-LSA(1s), if some ISL failures are not perceived by the source satellite, then the specified path may lead to a failed packet delivery.
Hence LiR-LSA(1s) achieves a lower packet delivery ratio.

\item Under OSPF-LSA(1s), if the link-state announcement is not in time, the forwarding decision made by intermediate nodes may not be the optimal one, which leads to a detour route.
Hence OSPF-LSA(1s) incurs a larger delay.
\end{itemize}

\textbf{Observation~3:}
Under LiR architecture, the two cases LiR-ODR and LiR-ODD can outperform LiR-LSA(1s) in terms of packet delivery ratio.
The main reason is that both LiR-ODR and LiR-ODD utilize the deterministic neighbor relation to handle the occasional ISL failures. 
However, LiR-LSA(1s) follows from the classic LSA designed for terrestrial Internet.
Such a scheme does not perform as well as ODD and ODR.

\begin{figure}
\centering
\subfigure[End-to-End packet delay]{\label{fig: Multicast_delay}
\includegraphics[width=0.88\linewidth]{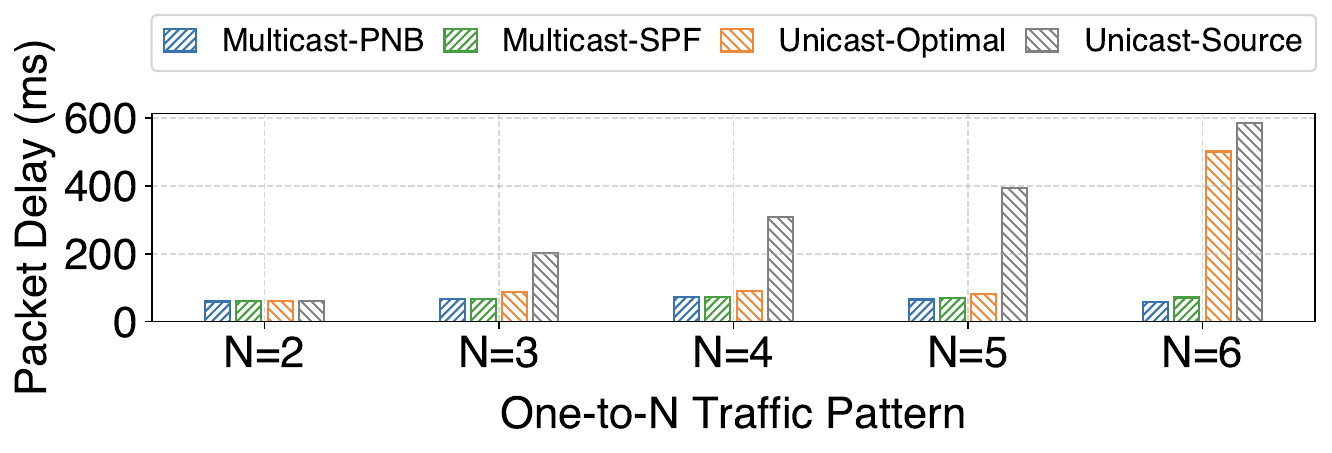}}
\subfigure[Packet delivery ratio]{\label{fig: Multicast_ratio}
\includegraphics[width=0.88\linewidth]{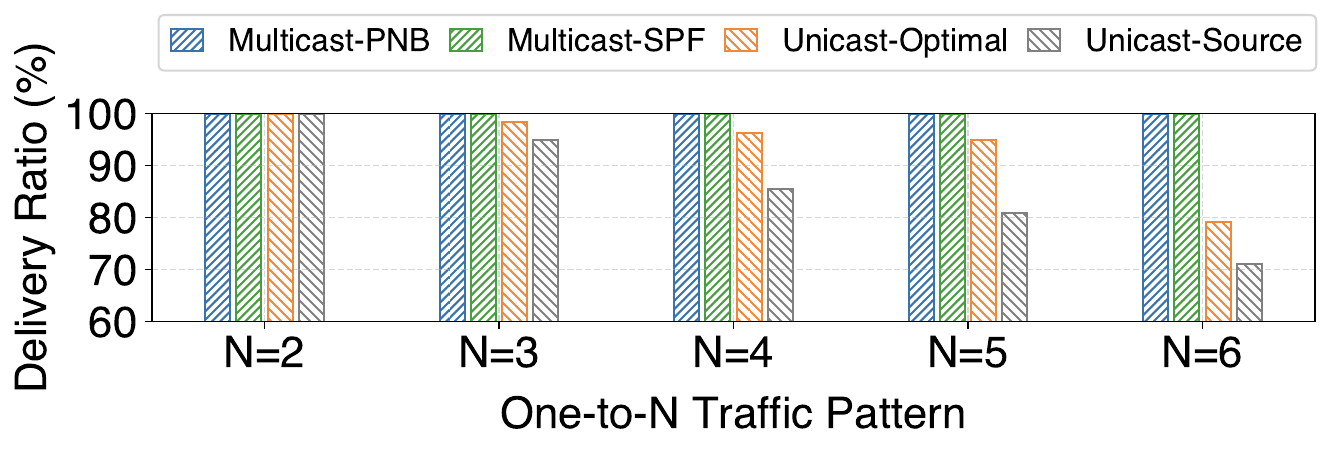}}
\caption{LiR architecture under one-to-many traffic pattern}
\label{fig: Multicast}
\end{figure}

\subsection{Multicast Transmission under LiR Architecture}
\label{Subsection: Multicast Transmission under LiR}

We evaluate the performance of our proposed LiR architecture under one-to-many traffic pattern.
Such a traffic pattern captures the scenario where the source satellite utilizes multicast to achieve seamless satellite-ground handover.
We consider multiple one-to-N source-destination pairs, i.e., $N\in\{2, 3, ..6\}$.
The sending rate of source satellites is 1.6~Mbps. 
Fig.~\ref{fig: Multicast} shows the results under different transmission pattern.
Each sub-figure consists of the following four cases:

\begin{itemize}

\item The case of Multicast-PNB (i.e., blue bars) corresponds to the primary-node-based multicast under LiR.

\item The case of Multicast-SPF (i.e., green bars)
corresponds to the shortest-path-first multicast under LiR.

\item The case of Unicast-Optimal (i.e., orange bars) corresponds to the optimal-encoding scheme under LiR.

\item The case of Unicast-Source (i.e., grey bars) corresponds to the source-encoding scheme under LiR.

\end{itemize}

We obtain the following observations from Fig.~\ref{fig: Multicast}.

\textbf{Observation~1:}
Comparing the blue/green bars to orange/grey bars in Fig.~\ref{fig: Multicast} shows that Multicast-PNB and Multicast-SPF outperform the other two cases in terms of end-to-end delay and packet delivery ratio.
This observation unveils the advantage of multicasting in LiR architecture.
Specifically, under the Multicast-PNB and Multicast-SPF cases, the source satellite encodes all the ISL identifiers in a single BF.
Although this may potentially lead to a high false positive rate, it significantly reduces the redundant packet delivery in a one-to-many traffic pattern.

\textbf{Observation~2:}
Comparing the green bars and blue bars reveals that the Multicast-PNB mechanism slightly outperforms the Multicast-SPF mechanism in terms of end-to-end delay.
This is because multicast-PNB mechanism incurs less incorrect forwarding.
For multicast-PNB, the link identifiers encoded in the in-packet BF consists of the path from the source to the primary node and from primary node to the other nodes.
For multicast-SPF, the link identifiers in the BF encompass the path from source to each node directly.
This approach may results in multiple paths from the source node to the same destination group are redundantly encoded into the BF.
Consequently, it leads to more incorrect forwarding and increased queuing delays.

\textbf{Observation~3:} 
Comparing the orange bars with the grey bars, we find that the case Unicast-Optimal outperforms Unicast-Source. 
The reason is similar to the observation obtained from Fig.~\ref{fig: multi-flow}.
That is, Unicast-Source leads to a higher false positive rate and generates more redundant forwarding, which ultimately results in a larger end-to-end delay and a lower packet delivery ratio.





\section{Conclusion and Future Work}
\label{Section: Conclusion and Future Works}
This paper proposes a Link-identified Routing (LiR) architecture for LEO satellite networks, which supports efficient source-route-style forwarding.
LiR identifies each ISL and enables satellites to encode ISL identifiers via the in-packet BF to specify the packet delivery path, which provides greater path representation efficiency than SRv6 and other techniques.
To mitigate the redundant forwarding caused by the false positive rate of the in-packet BF, we propose an optimal encoding policy which decreases the forwarding overhead.
Our experiment results show that the optimal encoding policy significantly outperforms the source encoding policy when BF size is small.
To address the intermittent ISLs, we design two on-demand ISL failure management schemes (i.e., LiR-ODD and LiR-ODR).

In the future, it would be interesting to investigate how to implement LiR architecture in Linux kernel.
Specifically, Linux kernel is based on the TCP/IP protocol stack.
A proper implementation should be incremental.

\bibliographystyle{IEEEtran}
\bibliography{ref}

\appendix
We introduce the difference between link-identified routing and node-identified routing.
To some degree, the node-identified method also fits the topology characteristics, and also allows us to represent an end-to-end path based on satellites (to be traversed) via the in-packet bloom filter (BF).
However, such a representation based on node-identified method \textbf{overlooks the direction information of the end-to-end path}.
This will cause many drawbacks.
In contrast, the link identifier used in this paper contains the direction information.
Hence a BF that records link identifiers can specify a concrete forwarding path.

\begin{figure}[H]
\centering
\subfigure[Link-identified routing]{\label{fig: multicast link identification}
\includegraphics[height=0.28\linewidth]{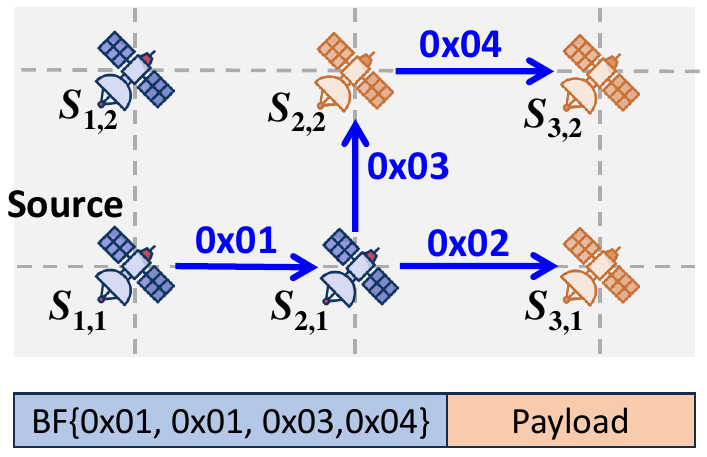}}\quad
\subfigure[Node-identified routing]{\label{fig: multicast node identification}
\includegraphics[height=0.28\linewidth]{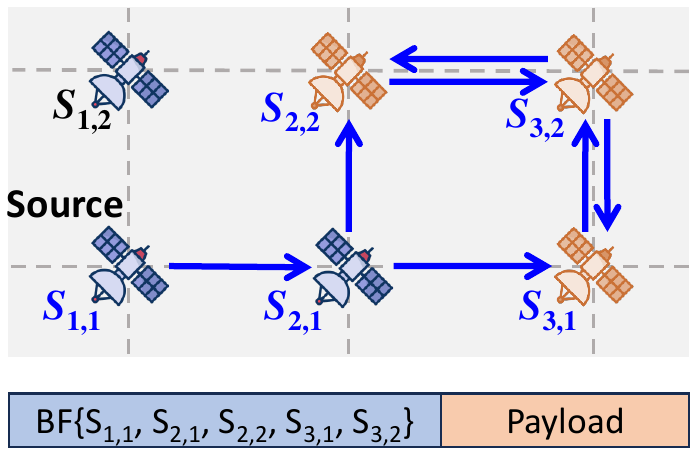}}
\caption{Multicast examples}
\label{fig: multicast under different identification}
\end{figure}

We illustrate the above difference based on Fig.~\ref{fig: multicast under different identification}.
In this example, satellite $\textit{S}_{1,1}$ multicasts to three orange satellites $\{\textit{S}_{2,2},\textit{S}_{3,2},\textit{S}_{3,1}\}$.
Fig.~\ref{fig: multicast link identification} illustrates how link-identified routing mechanism works.
Specifically, the source satellite $\textit{S}_{1,1}$ will encode four link identifiers \{0x01,0x02,0x03,0x04\} via the in-packet BF.
Such a BF can explicitly specify the multicasting tree, i.e., the blue arrows in Fig.~\ref{fig: multicast link identification}.
Fig.~\ref{fig: multicast node identification} illustrates how node-identified routing mechanism works.
Specifically, the source satellite $\textit{S}_{1,1}$ specifies the path via the in-packet BF based on the nodes $\{\textit{S}_{1,1},\textit{S}_{2,1},\textit{S}_{2,2},\textit{S}_{3,2},\textit{S}_{3,1}\}$.
For satellite $\textit{S}_{1,1}$, it checks whether its four neighbors are recorded in the packet, and then forwards the packet to satellite $\textit{S}_{2,1}$.
Then satellite $\textit{S}_{2,1}$ will find that both satellite $\textit{S}_{2,2}$ and satellite $\textit{S}_{3,1}$ are recorded in the packet, thus forwards the packet to both of them.
By this analogy, satellite $\textit{S}_{3,2}$ will receive duplicate packets from satellite $\textit{S}_{2,2}$ and satellite $\textit{S}_{3,1}$.
Even worse, satellite $\textit{S}_{3,2}$ may go on forwarding the packet from satellite $\textit{S}_{2,2}$ (or $\textit{S}_{3,1}$, respectively) to satellite $\textit{S}_{3,1}$ (or $\textit{S}_{2,2}$, respectively), forming a forwarding loop.
This phenomenon is due to the difference between link-specified path and node-specified path in multicast case.


\vspace{-20pt}

\begin{IEEEbiography}
[{\includegraphics[width=1in,height=1.25in,clip,keepaspectratio]{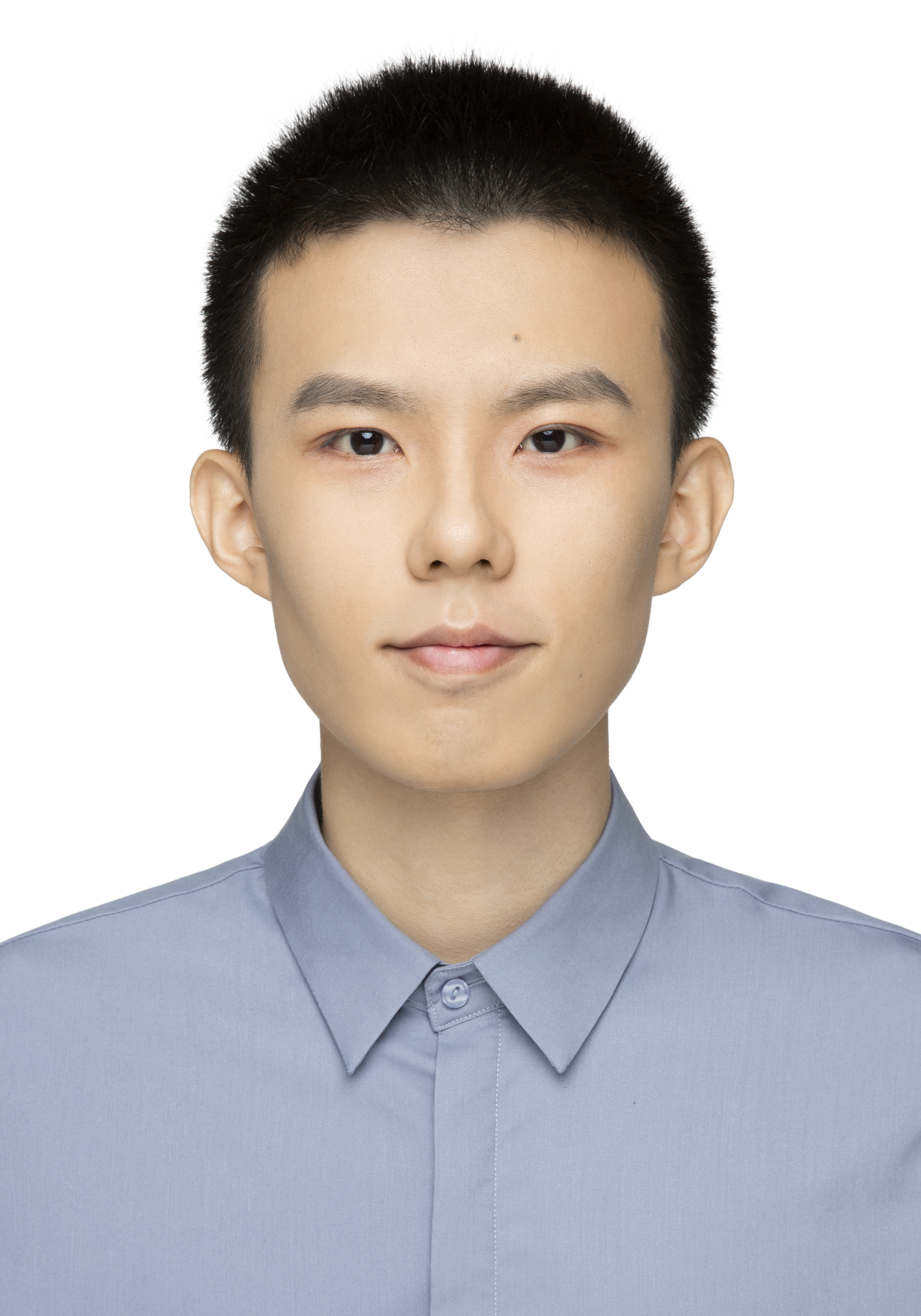}}] {Hefan Zhang} received the B.S. degree in School of Software Engineering from the Nanchang University, Nanchang, China, in 2022. 
He is currently pursuing the Ph.D. degree in the School of Computer Science and Engineering, Beihang University. His research interests include Internet architecture, Satellite networks routing, and Integration of satellite-terrestrial networks.
\end{IEEEbiography}

\vspace{-30pt}
\begin{IEEEbiography}
[{\includegraphics[width=1in,height=1.25in,clip,keepaspectratio]{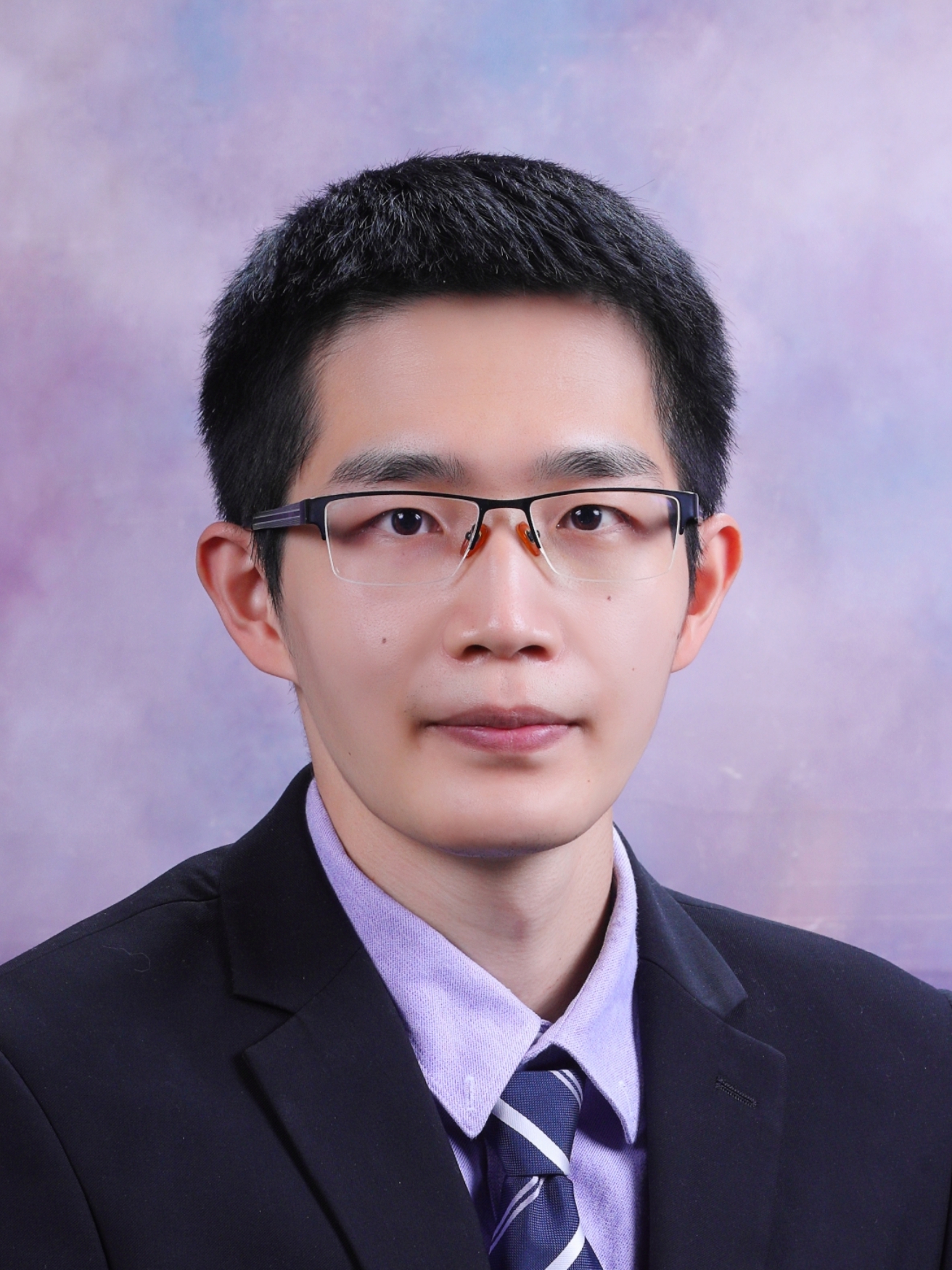}}]{Zhiyuan Wang} is an associate professor in School of Computer Science and Engineering, Beihang University.
He was a Post-Doctoral Fellow in Department of Computer Science and Engineering, The Chinese University of Hong Kong from 2019 to 2021.
He received his Ph.D. degree in Information Engineering, from The Chinese University of Hong Kong, in 2019. 
He received the B.Eng. degree in School of Information Science and  Engineering, from Southeast University, Nanjing, in 2016. 
His research interest includes network science.
\end{IEEEbiography}

\vspace{-30pt}
\begin{IEEEbiography}
[{\includegraphics[width=1in,height=1.25in,clip,keepaspectratio]{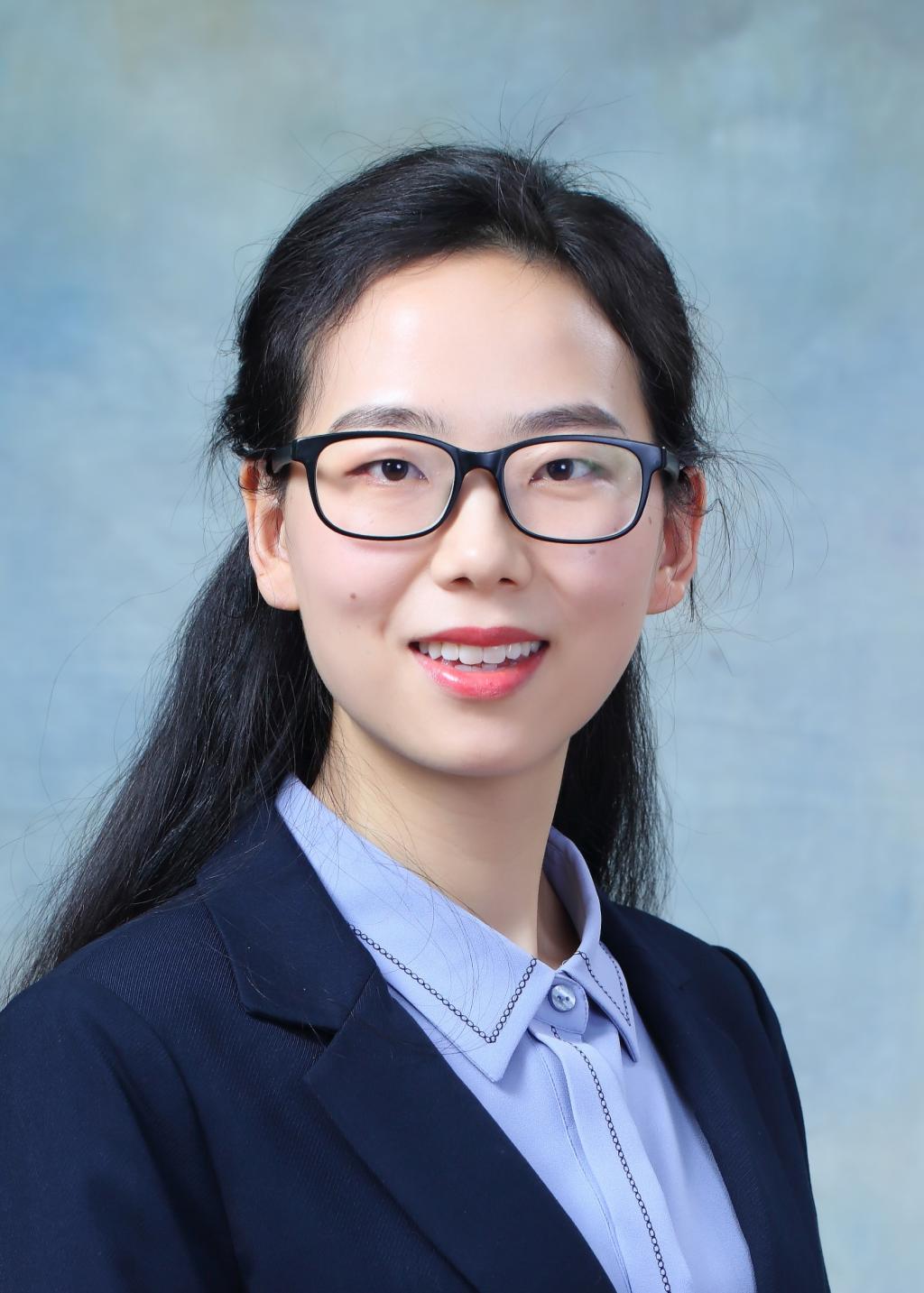}}]
{Shan Zhang} (Member, IEEE) received the Ph.D. degree in electronic engineering from Tsinghua University, Beijing, China, in 2016. She is currently an associate professor at the School of Computer Science and Engineering, Beihang University, Beijing, China. 
She was a postdoctoral fellow in the Department of Electronical and Computer Engineering, University of Waterloo, Ontario, Canada, from 2016 to 2017. 
Her research interests include mobile edge computing, wireless network virtualization, and intelligent management. 
\end{IEEEbiography}

\vspace{-30pt}
\begin{IEEEbiography}
[{\includegraphics[width=1in,height=1.25in,clip,keepaspectratio]{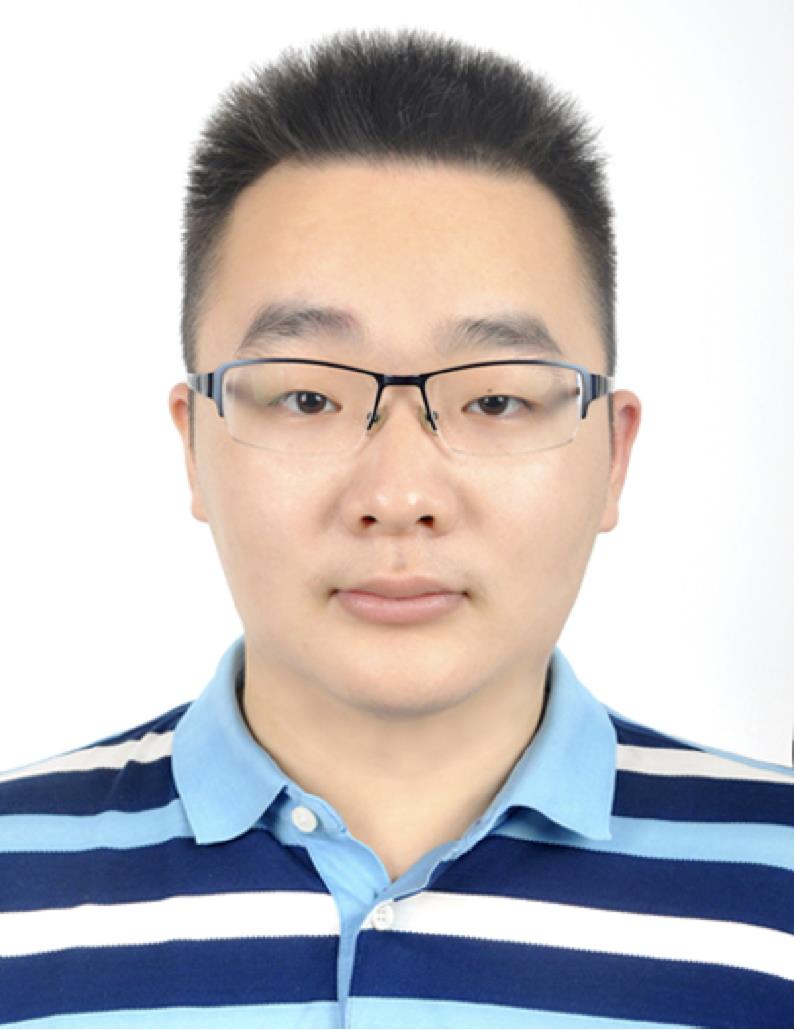}}]{Qingkai Meng} (Member, IEEE) received the Ph.D. degree in computer science and technology from Tsinghua University, Beijing, China, in 2022. 
He currently works at Beihang University as a post doctoral researcher. 
From September 2019 to October 2020, he was a visiting scholar at the Department of Computer Science, University of Wisconsin-Madison, funded by the Chinese Scholarship Council. 
His research interests include network architecture, datacenter network, transport protocol and programmable switch. 
\end{IEEEbiography}

\vspace{-30pt}
\begin{IEEEbiography}
[{\includegraphics[width=1in,height=1.25in,clip,keepaspectratio]{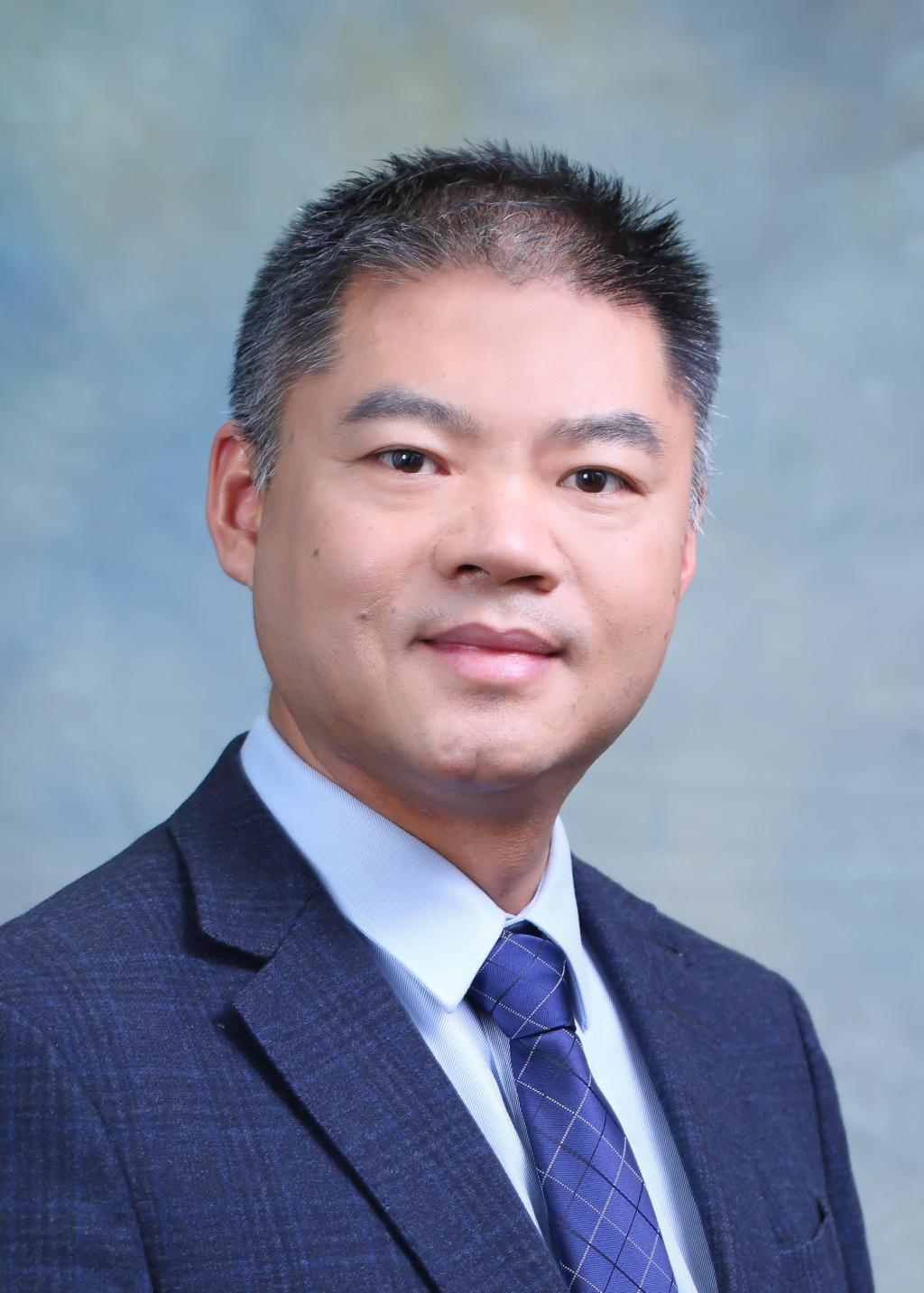}}]
{Hongbin Luo} (Member, IEEE) received the B.S. degree from Beihang University, in 1999, and the M.S. (with honors) and Ph.D. degrees in communications and information science from the University of Electronic Science and Technology of China (UESTC), in June 2004 and March 2007, respectively. 
He is currently a professor at the School of Computer Science and Engineering, Beihang University. 
From June 2007 to March 2017, he worked at the School of Electronic and Information Engineering, Beijing Jiaotong University.
His research interests include  network architecture, routing, and traffic engineering.
\end{IEEEbiography}

\end{document}